\documentclass[twocolumn,preprintnumbers,nofootinbib,prd,superscriptaddress,aps]{revtex4-1}

\usepackage[utf8]{inputenc} 
\usepackage{graphicx,amssymb,amsmath,amsthm,amsfonts,epstopdf,epsfig,epsf,times}
\usepackage[linktocpage]{hyperref}
\usepackage[usenames]{color}
\usepackage{epstopdf}

\graphicspath{{plots/}} 

\usepackage{textcomp}
\usepackage{bm}
\usepackage{dcolumn}
\usepackage{latexsym}
\usepackage{rotating}
\usepackage{hyperref}
\usepackage{color}
\usepackage{longtable}
\usepackage{enumerate}
\usepackage{tensor}
\usepackage{stmaryrd}
\usepackage[normalem]{ulem}

\usepackage{mathtools}
\usepackage{url}
\definecolor{coolblack}{rgb}{0.0, 0.18, 0.39}
\definecolor{darkred}{rgb}{0.5,0,0}
\definecolor{darkgreen}{rgb}{0,0.5,0}
\definecolor{darkblue}{rgb}{0,0,0.5}
\definecolor{lapislazuli}{rgb}{0.15, 0.38, 0.61}
\definecolor{venetianred}{rgb}{0.78, 0.03, 0.08}
\definecolor{bleudefrance}{rgb}{0.19, 0.55, 0.91}
\definecolor{dogwoodrose}{rgb}{0.84, 0.09, 0.41}
\definecolor{dogwoodrose}{rgb}{0.84, 0.09, 0.41}
\definecolor{darkorgane}{rgb}{1,0.549,0}
\hypersetup{colorlinks=true, citecolor=darkblue, linkcolor=darkblue, 
urlcolor = darkblue}

\definecolor{olive}{rgb}{0.5, 0.5, 0.0}

\renewcommand{\vec}[1]{\boldsymbol{#1}}

\newcommand{\ben}{\begin{enumerate}}
\newcommand{\een}{\end{enumerate}}

\newcommand{\leqsim}{\,\mbox{{\scriptsize $\stackrel{<}{\sim}$}}\,}

\newcommand{\del}{\partial}

\def\be{\begin{equation}}
\def\ee{\end{equation}}
\newcommand{\beq}{\begin{eqnarray}}
\newcommand{\eeq}{\end{eqnarray}} 
\newcommand{\ba}{\begin{align}}
\newcommand{\ea}{\end{align}}

\def\be{\begin{equation}}
\def\ee{\end{equation}}
	
\newcommand{\bea}{\begin{eqnarray}}
\newcommand{\eea}{\end{eqnarray}}

\begin{document}\title {\large Motion in time-periodic backgrounds\\
with applications to ultralight dark matter haloes at galactic centers}

\author{Mateja Bošković}
\affiliation{CENTRA, Departamento de F\'{\i}sica, Instituto Superior T\'ecnico -- IST, Universidade de Lisboa -- UL,
Avenida Rovisco Pais 1, 1049 Lisboa, Portugal}
\affiliation{Department of Astronomy, Petnica Science Center,
P. O. B. 6, 14104 Valjevo, Serbia}
\author{Francisco Duque}
\affiliation{CENTRA, Departamento de F\'{\i}sica, Instituto Superior T\'ecnico -- IST, Universidade de Lisboa -- UL,
Avenida Rovisco Pais 1, 1049 Lisboa, Portugal}
\author{Miguel C. Ferreira}
\affiliation{CENTRA, Departamento de F\'{\i}sica, Instituto Superior T\'ecnico -- IST, Universidade de Lisboa -- UL,
Avenida Rovisco Pais 1, 1049 Lisboa, Portugal}
\author{Filipe S. Miguel}
\affiliation{CENTRA, Departamento de F\'{\i}sica, Instituto Superior T\'ecnico -- IST, Universidade de Lisboa -- UL,
Avenida Rovisco Pais 1, 1049 Lisboa, Portugal}
\author{Vitor Cardoso}
\affiliation{CENTRA, Departamento de F\'{\i}sica, Instituto Superior T\'ecnico -- IST, Universidade de Lisboa -- UL,
Avenida Rovisco Pais 1, 1049 Lisboa, Portugal}
\affiliation{Perimeter Institute for Theoretical Physics, 31 Caroline Street North
Waterloo, Ontario N2L 2Y5, Canada}

\begin{abstract}
We consider motion in spherically symmetric but time-dependent backgrounds. This problem is of interest, for example, in the context of ultralight dark matter, where galactic haloes produce a time-dependent and periodic gravitational potential. We study the properties of motion of stars in such spacetimes, for different field strengths and frequency, and including dissipative effects. We show that orbital resonances may occur and that spectroscopic emission lines from stars in these geometries exhibit characteristic, periodic modulation patterns. In addition, we work out a fully relativistic {\it and} weak-field description of a special class of time-periodic geometries, that of scalar oscillatons. When applied to the galactic center, our results indicate that the motion of S2-like stars may carry distinguishable observational imprints of ultra-light dark matter.
\end{abstract}

\date{\today}

\maketitle

\tableofcontents

\section{Introduction}\label{sec:intro}

\subsection{Motion in General Relativity}

The problem of motion in General Relativity is a complex but fundamental one.
It is the motion of objects {\it and} of light that allows for precise
tests of the theory, by connecting it to observations. Conversely, the way that objects move allows one to infer, study and map the amount of matter contributing to the motion. When the object (a star, a planet, etc...) is idealized as point-like, it moves
-- to first approximation -- along geodesics of the spacetime ``generated'' by the rest of the universe.

Static, spherically symmetric objects in otherwise empty spacetime give rise to a Schwarzschild geometry. Geodesic motion around a Schwarzschild background has been studied for decades. The symmetries of the spacetime allow for three constants of motion, which simplify considerably the analysis and make the problem integrable. However, spherical symmetry does not necessarily imply staticity when matter pervades the geometry. For example, radially oscillating stars produce an effective geometry which is time-dependent in their interior but Birkhoff's theorem guarantees that the spacetime outside such a configuration is described by a Schwarzschild geometry \cite{bookWeinberg:1972}. 
Here, we do not have in mind the interior of stars, but rather the motion of planets, stars and compact objects within extended, dynamical matter distributions. One important example is described by a self-gravitating massive scalar, and is used to describe dark matter (DM), either at a fundamental level or as an effective description~\cite{Barack:2018yly}. The field equations describing a (real) minimally coupled massive scalar can give rise to spatially bound, time-dependent, spherically symmetric solutions~\cite{Bogolyubsky:1976nx,Copeland:1995fq,Honda:2001xg,Fodor:2006zs,Brito:2015yga}.


Thus, the question arises as to how matter moves in time-periodic geometries and whether the motion exhibits characteristic behavior.
Some aspects of this question were considered previously within a very specific context -- that of oscillating bosonic DM -- and within some approximations~\cite{Khmelnitsky:2013lxt,Blas:2016ddr,Ferreira:2017pth, AokiSoda:2017a}. 

Here we wish to consider the full problem of geodesic motion, in what looks like a classic problem in Newtonian physics and General Relativity: how do particles move in a time-dependent and periodic gravitational potential? 

The general description and discussion of symmetries of motion of massive particles is provided in Section~\ref{sec:motionformalism}. In Section~\ref{sec:examples}, we construct and examine, both analytically and numerically, several examples. Our focus is mostly on whether and how resonances and instabilities manifest in this motion, as they are the most distinct signatures of time-periodic backgrounds.

\subsection{Galactic haloes}

Although extremely successful on a large variety of scales, the cold dark matter (CDM) model is met with several problems at galactic scales. 
One way to alleviate these problems is to rely on baryonic feedback processes\footnote{Such as reionization, (baryonic) radiative cooling, star formation, supernovae explosions etc.}. However, these mechanisms aren't at present consistently and uniformly successful in solving small scale challenges. Other approaches, within the DM paradigm, rely on modifying linear or non-linear aspects of the structure formation. These include modification of very slow small-scale suppression in the primordial power spectrum of the DM density fluctuations and/or importance of DM self-interaction~\cite{Bullock:2017xww}. The first approach is natural in the warm DM model, where DM particles have significant thermal velocities, and in the case of ultra-light axions (ULA), where the power spectrum small scale suppression occurs because of the large particle's de Broglie wavelength~\cite{Hui:2016ltb, Marsh:2016rep}. 
It should be noted that there is no a priori reason why these alternative models should solve all of the small scale problems (``catch all'' solution), while reproducing large-scale success of the CDM, as some (or all) of them may indeed be solved by progress in the understanding of baryonic feedback (however, see Ref. \cite{Famaey:2017xou}). It is, thus, necessary to think about direct tests of such models. Both warm DM and ULA have been also considered in the context of mixed DM cosmologies \cite{AnderhaldenDiemand:2013, MarshSilk:2013}.

When the ULA particle has a mass in the range $10^{-21}\leqsim m[\text{eV}] \leqsim 10^{-23}$ and a weak self-interaction\footnote{For scalar DM models with strong self-interactions see Appendix \ref{AppGPP}.}, it is a candidate for the dominant component of the DM and is usually referred to as a fuzzy DM (FDM).
ULA particles are motivated by QCD and string theory~\cite{Marsh:2017hbv} and are described by a real scalar field, subject of this work. Solutions to the field equations should, in this context, be interpreted as describing a region of the dark halo. Because the system is dilute, it can be analysed in the weak field limit, along with the study of test particle motion inside such a background. We discuss and study this scenario in Section~\ref{sec:NOscillatons},  Section~\ref{sec:darkhalo} and Appendices \ref{AppEliptical} and \ref{AppBinary} and show that observable consequences of motion in ULA background may be found in the analysis of motion of S2-like stars, at the Galactic center.

\subsection{Compact objects and BH mimickers}
In the other limit, compact configurations made from a real scalar field, dubbed oscillatons, and motion inside them, e.g. extreme mass ratio inspirals (EMRI), can be of relevance in understanding the nature of compact, massive and dark objects~\cite{Cardoso:2017njb,Macedo:2013qea,Macedo:2013jja,Ferreira:2017pth,HelferLim:2018, Barack:2018yly}. Thus, our results can also make close contact with attempts at explaining observations of compact objects. 

We review the description of oscillatons in both the strong-field (numerically, in Section~\ref{sec:ROscillatons}) and the weak-field regime (analytically, in Section~\ref{sec:NOscillatons} and Appendices \ref{AppEKGNewt} and \ref{AppGPP}). We also discuss how dynamical oscillatons spacetimes are. In Section \ref{sec:applications_oscillatons} we confirm previous results~\cite{2006GReGr..38..633B} for the motion in oscillatons, show that no resonances can be excited by motion in them and develop an analytical framework for description of motion in dilute oscillatons. The motion in self-interacting Newtonian oscillatons is briefly discussed in Appendix~\ref{AppGPP_self_int_motion}. The motion of light in oscillaton spacetimes, in the context of gravitational redshift, is discussed in Appendix~\ref{AppGredshift}.

\section{Setup}
\subsection{Dynamic, time-periodic spacetimes} \label{sec:spacetime}
We adopt geometric units ($G=c=1$) and consider a spherically symmetric (Lorentzian) spacetime which in radial coordinates $(t,r,\vartheta,\varphi)$ takes the form
\begin{equation}
ds^2=g_{\mu\nu}d x^\mu dx^\nu=-Adt^2+Bdr^2+r^2d\Omega^2\,,\label{eq:metric}
\end{equation}
where $A\equiv A(t,r)>0,\, B\equiv B(t,r)>0$ and $d\Omega^2=d\vartheta^2+\sin^2(\vartheta)d\varphi^2$ is the metric on the two-sphere.
The functions $A$ and $B$ are taken to be periodic functions of the time coordinate.

In most of this work, the spacetime is to a good approximation static, and the time-dependence is but a small deviation away from staticity. 
Thus, we find it convenient to expand the metric around a static reference spacetime $g^{(0)}_{\mu\nu}$
\be
g_{\mu\nu}=g^{(0)}_{\mu\nu}+\epsilon g^{(1)}_{\mu\nu}\,.\label{metric_expansion}
\ee
Having in mind specific applications, we consider spacetimes for which
\beq
A(t,r)&=&a_0(r)+\epsilon a_1(r)\cos\left(2\omega t\right)\,,\label{eq:weakly1}\\
B(t,r)&=&b_0(r)+\epsilon b_1(r)\cos\left(2\omega t\right)\,,\label{eq:weakly2}
\eeq
thus describing time-periodic geometries with period $T=\pi/\omega$.

\subsection{An example: oscillatons} \label{sec:Oscillatons}

One  possible time-periodic metric is obtained in the context of minimally coupled scalar fields, giving rise to 
self-gravitating structures~\cite{Seidel:1991zh,Brito:2015yga}.
These are time-dependent, spherically symmetric, real~\footnote{Complex scalar field counterparts to oscillatons are known as boson stars, whose metric is stationary but have an harmonic boson~\cite{Liebling:2012fv}. This configurations have $\text{U}(1)$ symmetry and are hence protected by a charge.} scalar field solutions $\Phi(t,r)$ of the coupled Einstein-Klein-Gordon (EKG) equations derived from the action
\begin{equation}
\label{eq:action}
S = \int d^4x \sqrt{-g} \left(\frac{1}{16 \pi} R - \frac{1}{2} g^{\alpha \beta}\partial_{\alpha}\Phi \partial_{\beta} \Phi - U(\Phi)\right)\,,
\end{equation}
where  $U(\Phi)$ is the scalar potential. 
The canonical axion potential is
$U(\Phi) = \mu^2 f_{a}^2 (1-\cos (\Phi/f_{a} ))$, and the object is then regarded as an axion star~\cite{Helfer:2017a}. Here,
$f_{a}$ is relevant energy scale (decay constant), $\mu = m/\hbar$ is the mass parameter and $m$ is the mass of the scalar field. 

If the axion potential is expanded in a Taylor series for small scalars, and the field is interpreted in a classical sense, we call such an object oscillaton. Unless stated otherwise, we focus on non-self-interacting oscillatons with the scalar potential containing only the mass term
\begin{equation}
U(\Phi) = \frac{1}{2} \mu^2 \Phi^2\,.
\end{equation}
The dynamics and stability of these objects were studied in Refs.~\cite{Seidel:1991zh,Alcubierre:2003sx,Okawa:2013jba,Brito:2015yfh}, where a set of stable ground states were found (excited states are unstable and we do not discuss them here). These solutions actually have a small radiating tail~\cite{Grandclement:2011wz,Page:2003rd,Fodor:2009kg}, but the mass-loss rate is for much of the parameter space larger than a Hubble time. Such solutions can be formed through gravitational collapse and cooling mechanisms~\cite{Seidel:1993zk,Guzman:2006yc,Guzman:2004wj,Okawa:2013jba,Brito:2015yfh}.

By minimizing action in Eq.~\eqref{eq:action} we obtain:
\beq
&&R_{\alpha \beta }-\frac{R}{2}g_{\alpha \beta }= 8\pi  T_{\alpha \beta }\,,\label{eq:einstein}\\
&&T_{\alpha \beta} = \partial_{\alpha} \Phi \partial_{\beta} \Phi - \frac{1}{2} g_{\alpha \beta} \left( g^{\gamma \sigma} \partial_{\gamma}\Phi \partial_{\sigma}\Phi + \mu^2 \Phi^2 \right)\,,\label{eq:stressenergy}\\
&&\frac{1}{\sqrt{-g}} \partial_{\alpha} \left( \sqrt{-g} g^{\alpha \beta} \partial_{\beta} \Phi \right) = \mu^2 \Phi\,. \label{eq:kg}
\eeq
From now on, and for numerical convenience, we will be writing the spherically symmetric metric as
\begin{equation}
 ds^2 = B(t,r) \left(- \frac{1}{C(t,r)}dt^2 +  dr^2 \right) + r^2 d\Omega^2 \label{eq:metric2}\,,
\end{equation}
set units such that $\mu = 1$ and redefine the scalar through
\be
\Phi = \frac{\Psi}{\sqrt{4 \pi}}\,.
\ee

With these definitions, Eqs.~\eqref{eq:einstein} and \eqref{eq:kg} lead to:
\begin{align}
  & -\frac{B'}{r B}+B \left(\Psi^2-\frac{1}{r^2}\right)+C \partial_t{\Psi}^2+\Psi'^2+\frac{1}{r^2} = 0\,, \label{eq:1of4}\\
  & 2 \Psi' \partial_t{\Psi}-\frac{\dot{B}}{r B} = 0\,, \label{eq:2of4}\\
  & \frac{B'}{B}+B \left(r \Psi^2-\frac{1}{r}\right)+\frac{1}{r} = \frac{C'}{C}+r C\partial_t{\Psi}^2+r \Psi'^2\,, \label{eq:3of4}\\
  & r B \Psi+\frac{r C'}{2C} \Psi'+\frac{r \partial_t{C}}{2} \partial_t{\Psi}-2\Psi'-r \Psi''+r C \partial^2_t{\Psi} = 0\,. \label{eq:4of4}
\end{align}
Here and throughout, primes stand for radial derivatives while $\partial_t\equiv \partial/\partial t$ and dot for proper time derivatives.

\subsubsection{Fully relativistic results} \label{sec:ROscillatons}
Oscillaton geometries can be obtained through the expansion:
\begin{align} 
  & B(t,r) = \sum_{j=0}^{\infty} b_{j}(r) \cos(2 j \omega t)\,, \label{EKGmetricB} \\
  & C(t,r) =  \sum_{j=0}^{\infty} c_{j}(r) \cos(2 j \omega t)\,, \label{EKGmetricC} \\
  & \Psi(t,r) =  \sum_{j=0}^{\infty} \Psi_{j+1}(r) \cos([2 j+1] \omega t)\,, \label{EKGmetricPhi}
\end{align}
truncated at a finite $j$, which depends on the accuracy necessary (for most cases $j=1$ is accurate at a $\sim 1\%$ level).
Accordingly, we will also use reference spacetimes for which,
\begin{align}
  & B(t,r) =  b_{0}(r) + b_{1}(r) \cos(2 \omega t)\,,\label{eq:expB}\\
  & C(t,r) = c_{0}(r) + c_{1}(r) \cos(2 \omega t)\,,\label{eq:expC}\\
  & \Psi(t,r) = \Psi_{1}(r) \cos(\omega t) + \Psi_{2}(r) \cos(3 \omega t)\,.\label{eq:expPhi}
\end{align}

\begin{figure}
\centering
\includegraphics[width=\columnwidth]{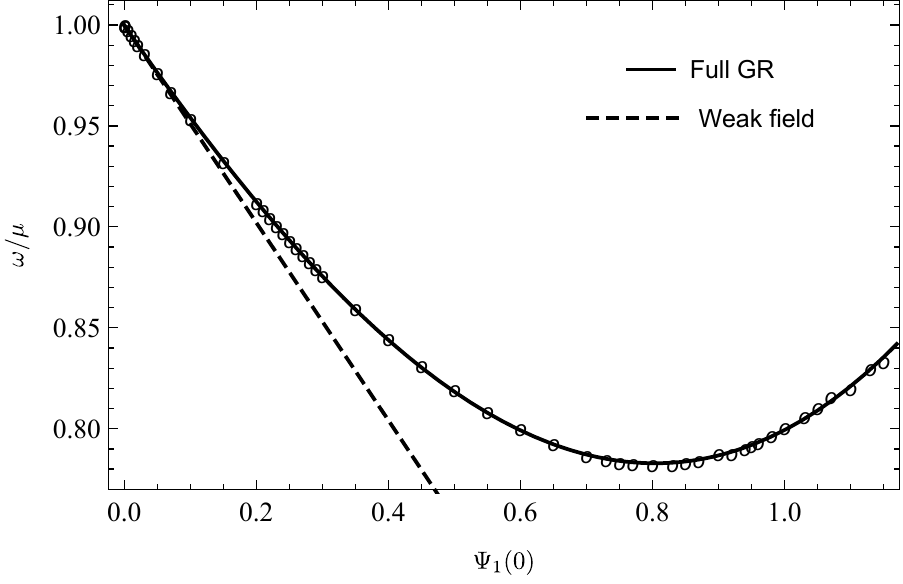}
\caption{The fundamental frequency $\omega$ of a scalar oscillaton, as a function of the central value of the scalar field. The minimum value of the frequency is given by $\omega/\mu \sim 0.782$.
The dashed line is the weak-field prediction, discussed in the next section, and agrees well with the full relativistic results at low compactness.}
\label{fig:OvsPhi0}
\end{figure}

The coefficients $b_{0},\,b_{1}...$ can be obtained by inserting the expansion above in the equations of motion, and requiring the vanishing of each harmonic term, order by order.
In this particular case, one finds six ODEs for the variables ${b_0,b_1,c_0,c_1,\Psi_1,\Psi_2}$. These equations are solved numerically using a shooting method; we have the freedom to choose the central value of $\Psi_1$, but we fix the radial metric functions, so that $b_0(0) = 1$ and $b_1(0) = 0$ and the derivatives of the scalar field components $\Psi_1'(0) = \Psi_2'(0) = 0$. Finally, we impose asymptotic flatness, requiring that as $r \to \infty$ the functions $\Psi_1,\Psi_2, b_1, c_1 \to 0$ and $c_0,b_0 \to 1$. In the end of the process we obtain, for each value of the central magnitude of the scalar field, the profiles of all the components of the metric and the scalar field as well as a value for the fundamental frequency $\omega$ of the oscillaton -- see Fig.~\ref{fig:OvsPhi0}.
\begin{figure}
\centering
\includegraphics[width=\columnwidth]{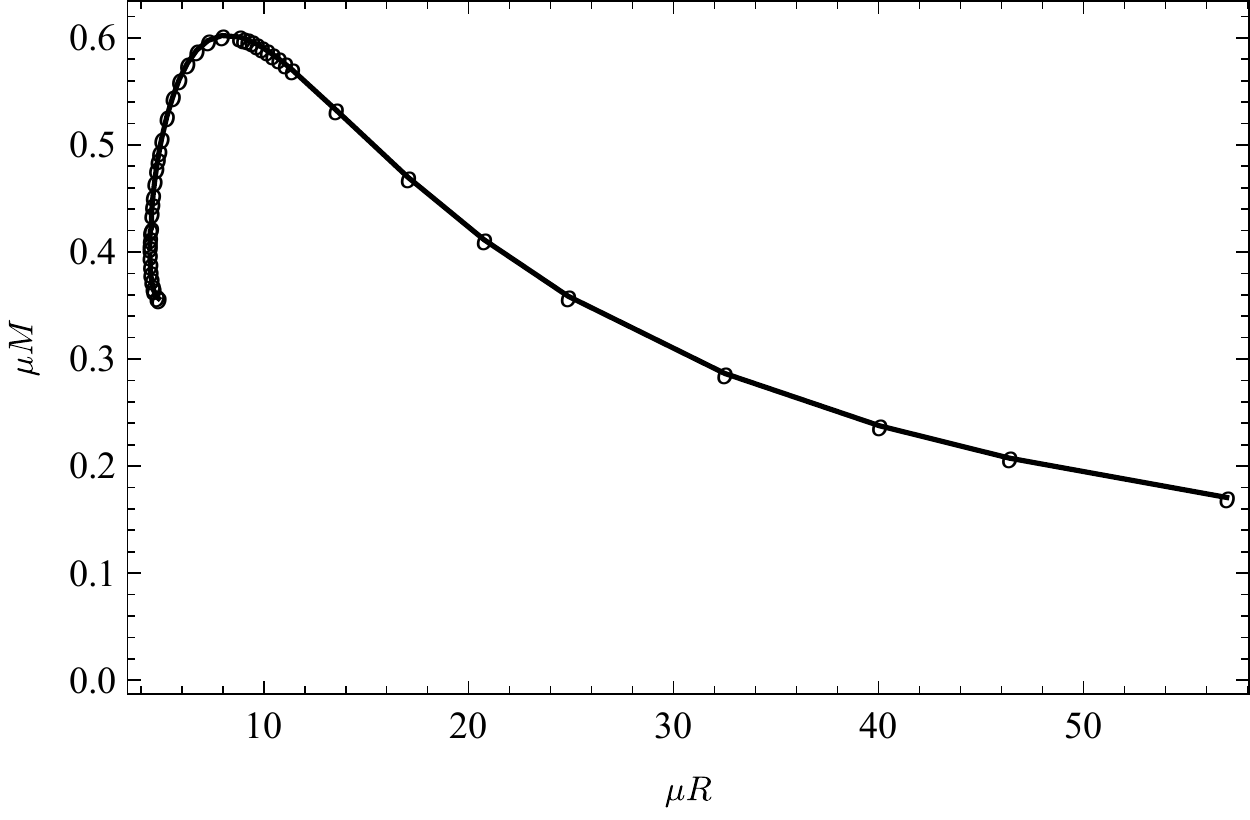}
\caption{Mass of the scalar oscillatons as a function of its radius. The maximum mass of stable configurations is $\mu M_{\text{max}} \sim 0.6$ and corresponds to a compactness $\mathcal{C} \sim 0.07$. These values mark the boundary between stable and unstable oscillaton configurations \cite{UrenaLopez:2002gx}: oscillatons with larger radii are stable, and are unstable for smaller radii. It is in the unstable branch of this plot that the maximum compactness is attained, $\mathcal{C}_{\text{max}} \sim 0.1$.}
\label{fig:MvsR}
\end{figure}
Notice that since $A(t,r) = B(t,r)/C(t,r)$ [see Eqs.~\eqref{eq:metric} and \eqref{eq:metric2}] , the coefficients of $A$ are obtained like this
\beq
a_0&=& \frac{2 b_0 c_0- b_1 c_1}{2 c_0^2+c_1^2}\,,\\
a_1&=& \frac{2 b_1 c_0 - 2 b_0 c_1}{2c_0^2 - c_1^2}\,,
\eeq
such that $A$ is written as
\be
 A(t,r)\equiv \frac{B}{C} = a_0(r) + a_{1}(r) \cos(2 \omega t)\,.\label{EKGmetricA}
\ee
Given that the solutions are spherically symmetric and asymptotically flat, the effective mass of these configurations is given by the following expression~\footnote{In this expression we recover the reduced mass $\mu$ of the scalar field.}
\be 
M=\frac{1}{\mu} \lim_{r \to \infty}\frac{r}{2}\left(1 - \frac{1}{b_{0}(r)}\right)\,.\label{eq:OscillatonMass}
\ee
We (arbitrarily) define the radius of the oscillaton as the location at which $98\%$ of the total mass is contained. Using these definitions, we obtain a good agreement with previous works on the subject -- see Figs.~\ref{fig:OvsPhi0}-\ref{fig:MvsR} and compare with Refs.~\cite{Seidel:1991zh,UrenaLopez:2001tw,Brito:2015yfh}.

The dynamical oscillaton spacetime can be characterized by comparing the magnitude of its time-dependent to its time-independent components.
These quantities depend on the compactness $\mathcal{C}$ of the spacetime,
\be
\mathcal{C} = \frac{M}{R}\,.
\ee

Our numerical results indicate that at small $\mathcal{C}$, and restoring the mass $\mu$, one has
\be
\mu R \approx \frac{9.8697}{\mu M}\,.\label{eq_C_num}
\ee

At large distances, the scalar profile decays exponentially and the spacetime is described by the Schwarzschild geometry. We thus focus on the metric components close to the origin, $r\ll 1/(M\mu^2)$. Our numerical results, for $\mathcal{C} < 0.07$, are described by:
\beq
\frac{a_1(0)}{a_0(0)}       &\sim&  6.2 \mathcal{C}+21.8 \mathcal{C}^2 -126 \mathcal{C}^3 + 6160.2 \mathcal{C}^4\,,\\
\frac{|b_1(0.5)|}{b_0(0.5)} &\sim& -0.0003\mathcal{C}+ 0.08 \mathcal{C}^2 - 6.3 \mathcal{C}^3 +325.8\mathcal{C}^4  \,.
\eeq
The error associated is of order $0.3\%$ for $a_1/a_0$ and $2\%$ for $b_1/b_0$.
From these fits, we see that the time-dependent part of the $g_{tt}$ component isn't always subdominant with respect to the corresponding static part. Unlike the time-dependent part of $g_{rr}$, which remains subdominant for all oscillatons, we see that for $g_{tt}$ the time-dependent part grows such that its magnitude becomes comparable, and even dominant, to the magnitude of the static part. 

One can take a closer look at the way in which compactness influences the spacetime metric by observing that its components can be written, for $r\mu < 1$ and $\mathcal{C} < 0.01$, as:
\beq
\label{eq:EKGmetricCRexpansion}
  a_0(r) &=& f_{a0}(\mathcal{C}) + g_{a0}(\mathcal{C})\mu^2 r^2\\
  a_1(r) &=& f_{a1}(\mathcal{C}) + g_{a1}(\mathcal{C})\mu^2 r^2 \\
  b_0(r) &=& f_{b0}(\mathcal{C}) + g_{b0}(\mathcal{C})\mu^2 r^2 \\
  b_1(r) &=& f_{b1}(\mathcal{C}) + g_{b1}(\mathcal{C}) \mu^2 r^2
\eeq
where the coefficients depend only on the compactness and are given in Table~\ref{tab:funcs}. We have also restored the mass $\mu$ for clarity.
The errors on the corresponding functions, in this range of values, are at most of $(0.6,3.4,0.05,3.0)\%$ for $a_0,\,a_1,\,b_0,\,b_1$ respectively.

\begin{table}[]
\centering
\caption{The behavior of oscillaton spacetimes at small radii, as described by \eqref{eq:EKGmetricCRexpansion}, where $\mathcal{C} \equiv M/R$. These results were obtained for $\mathcal{C} < 0.01$.}
\label{tab:funcs}
\begin{tabular}{ll}
$(f_{a0},\,f_{a1})=$ & \!\!\!\!\!\!\!($1 - 6.456 \mathcal{C} - 1673.7 \mathcal{C}^3\,,\quad6.163 \mathcal{C} - 1400.5 \mathcal{C}^3$)\\
$(g_{a0},\,g_{a1})=$ & \!\!\!\!\!\!\!($1.808 \mathcal{C}^2 + 77.162 \mathcal{C}^3\,,\quad -5.486 \mathcal{C}^2 - 2.820 \mathcal{C}^3$)\\
\,\,\,\,\,\qquad$f_{b0}=$ &  \!\!\!\!\!$1 - 0.819 \mathcal{C}^3 + 156.20 \mathcal{C}^4 - 6107.0 \mathcal{C}^5$\\
\,\,\,\,\,\qquad$g_{b0}=$ &  \!\!\!\!\!$1434.94 \mathcal{C}^3 - 163338 \mathcal{C}^4 + 5.816 \mathcal{C}^5$\\
\,\,\,\,\,\qquad$f_{b1}=$ &  \!\!\!\!\!$0.013 \mathcal{C}^3 - 2.34 \mathcal{C}^4 + 64.604 \mathcal{C}^5$\\
\,\,\,\,\,\qquad$g_{b1}=$ &  \!\!\!\!\!$-5.56 \mathcal{C}^3 - 63.71 \mathcal{C}^4 - 976.58 \mathcal{C}^5$
\end{tabular}
\end{table}
\subsubsection{Newtonian regime} \label{sec:NOscillatons}
The previous results focused on numerical solutions of the field equations. We now wish to make contact with semi-analytical results, valid in a ``Newtonian'' regime
when the velocities and densities involved are small~\cite{Seidel:1990jh,UrenaLopez:2002gx,Guzman:2004wj}. In this regime, we assume (something borne out of numerical calculations, see Fig.~\ref{fig:OvsPhi0}) that the frequency of the scalar field $\omega\sim \mu$. To be concrete, we take $\omega^2=\mu^2+k^2$ so that, up to the second order in wave number $k$, we can write
\begin{equation} \label{eq:Nomega}
\omega=\mu+\frac{k^2}{2\mu}+{\cal O}(k^4)\,.
\end{equation}
The group velocity of the field is equal, up to the first order in $k$, to $v=k/\mu$ and is our expansion parameter. As the field is ``trapped'' by self-gravity, $k^{2}<0$ and we expect for the long-range behaviour to be of the form $\psi \sim e^{i k r} \sim e^{-|k| r}$.

For the details of the expansion reader is referred to the Appendix \ref{AppEKGNewt} and we only cite here the form of metric coefficients and the field ansatz:
\beq
A(t,r)&=& 1 + 2V(r)+ 2V_{1}(r) \cos(2 \omega t)+{\cal O}(v^4)\,,\label{eq:NexpA}\\
B(t,r)&=& 1 + {\cal O}(v^4)\,,\label{eq:NexpB}\\
\Psi(t,r) &=& \psi(r) \cos(\omega t)\,.\label{eq:NexpPhi}
\eeq
We will from now on set $\mu=1$ and refer to $V(r)$ as the Newtonian potential and to $V_{1}(r)\cos(2\omega t)$ as the time-dependent potential. Working consistently up to second order in $v$, EKG system reduces to:
\beq
e\psi&=&-\frac{1}{2r}(r\psi)''+V\psi\,,\label{eq:NSchrodinger}\\
(rV)''&=&\frac{1}{2} r\psi^2\,,\label{eq:NPoisson}\\
V'_{1}&=&-\frac{1}{2} r\psi^{2}\,,\label{eq:NV2}
\eeq
where $e=k^2/2<0$. Note that the equations \eqref{eq:NSchrodinger} and \eqref{eq:NPoisson} are decoupled from \eqref{eq:NV2} and form the Schrödinger-Poisson (SP) system~\cite{Hui:2016ltb, Guzman:2004wj}. When the additional self-interacting potential is present this system is called Gross-Pitaevskii-Poisson  system \cite{Chavanis:2011} (see Appendix \ref{AppGPP}). 
Equation \eqref{eq:NV2} is present for oscillatons (and not for boson stars where the field is complex and harmonic) and is responsible for the time-dependence of the $A(t,r)$ metric coefficient. We can find the mass of the Newtonian oscillaton as $M=\int_{0}^{\infty} d V \rho(r)$, where $\rho(r) \equiv T_{tt}=\psi^2/(8\pi)$, and see that by definition it does not depend on the function $b_0(r)$ as is the case in general (\ref{eq:OscillatonMass}) and as expected from fully relativistic analysis (see Figure 1 in Ref.~\cite{Brito:2015yfh}).

Analytical solutions for these systems in general do not exist but there is a high precision approximate analytical solution 
in the case of the non-self-interacting fields~\cite{KlingRajaraman:2017}, which is the focus of this work.
Non-self-interacting oscillatons exhibit a Yukawa-like behavior at large distances. Thus, there is no well-defined notion of surface, even at a Newtonian level. The radius of this kind of object is defined as we did in the fully relativistic case. 

As the SP system admits scale symmetry, solutions corresponding to different masses can be obtained from a unique solution by rescaling~\cite{Guzman:2004wj,KlingRajaraman:2017}. 
The scaling that leaves SP system invariant for various parameters is given by,
\beq 
r    &\to& \frac{r}{\lambda}\,,\,e \to \lambda^2 e\,,\nonumber\\
\psi &\to& \lambda^2 \psi\,,\,V \to \lambda^2 V\,,\, V_{1} \to \lambda^2 V_{1}\,,\, M  \to \lambda M\,,\label{eq:NSPscale}
\eeq
where $\lambda$ is the scale factor. We will fix this factor as in Ref.~\cite{KlingRajaraman:2017} by identifying $2\lambda^2=-e$. A scale-independent field is found by expanding field around zero value of the radial coordinate and at infinity and matching these solutions. The free parameters are found by fitting this solution onto the numerical solution of the scale invariant SP system. These parameters are proportional to the scale invariant value of the central $(s_{0})$ and long-range $(\alpha)$ field expansion, the scale invariant mass $(\beta)$ and linearly related to the central value of the scale-invariant Newtonian potential $(v_{0})$. Technical details are left to Appendix \ref{AppGPP_non_self}. The numerical values of these parameters, along with the scale invariant radius $Z$, are:
\beq
s_{0} &=&  1.022\,,\quad v_{0} =  0.938 \nonumber\,,\\
\alpha &=& 3.495\,,\quad \beta =  1.753\nonumber\,,\\
Z  &= &  5.172\,.
\eeq
From Eq.~\eqref{eq:NSPscale}, it is obvious that the scaling between mass and radius is of the form
\begin{equation} \label{eq:R_M_N_Osc}
R = \frac{Z\beta}{M}=\frac{9.065}{M}
\end{equation}
and $\lambda=\sqrt{\mathcal{C} Z/\beta}$. Notice the excellent agreement with the low compactness full numerical result, 
Eq.~\eqref{eq_C_num}. From the scaling relations, we can find the dependence of the field frequency \eqref{eq:Nomega} on the central value of the field
\begin{equation}
\omega=1-\frac{\psi(0)}{2s_{0}}.
\end{equation}
The plot of this function is superposed on the relativistic $\omega-\Psi(0)$ plot  (Fig. \ref{fig:OvsPhi0}). We can see that the agreement for small values of $\Psi(0)$ is very good.


We will now provide comparison between small radius metric coefficients expansion in terms of compactness $\mathcal{C}$ obtained in fully relativistic analysis summarized in \eqref{eq:EKGmetricCRexpansion} and Table \ref{tab:funcs} and in Newtonian limit. The small $r$ behaviour of Newtonian oscillaton density is (see Appendix~\ref{AppGPP_non_self})
\begin{equation} \label{eq:Ndensitysmallr}
\rho(r) = \Lambda (a+b(\lambda r)^2) + {\cal O}(r^4)\,,
\end{equation}
where $\Lambda=\lambda^4/8\pi $, $a=s_{0}^2,\,b=-s_{0}^2 v_{0}/3$.

Newtonian non-self-interacting oscillatons do not have defined surface and the normalisation procedure for the Newtonian potential is not the same as for the sphere in Newtonian gravity. We have
\begin{equation}
V(r) = -\int^{\infty}_{0}\frac{dr}{r^2}m(r)+\int^{r}_{0}\frac{dy}{y^2}m(y)\,,
\end{equation}
where $m(r)$ is the Newtonian mass function.
The first term -- proportional to $\mathcal{C}$ (as can be seen from a dimensional analysis) -- is integrated using the full expansion described in the Appendix~\ref{AppGPP_non_self}. The second term reduces to $2\pi \Lambda a r^2/3$ at ${\cal O}(r^3)$. Similarly
\begin{equation}
V_{1}(r) =4\pi \int^{\infty}_{0}dr r\rho (r)-4\pi \int^{r}_{0}dy  y \rho(y).
\end{equation}
The small-$r$ expansion for the second term gives us $-2\pi\Lambda a r^2$  at ${\cal O}(r^3)$. The first, of the order $\mathcal{C}$, is integrated using the full expansion. We get the following results for the parameters defined in \eqref{eq:EKGmetricCRexpansion},
\beq \label{eq:Nmetricexpansion}
f_{a0}&=& 1 - 5.720 \mathcal{C}\,,\quad f_{a1}= 5.720 \mathcal{C}\,,\\
g_{a0}&=& 1.514 \mathcal{C}^2\,,\quad g_{a1}= - 4.543 \mathcal{C}^2\,,\\
f_{b0} &=& 1\,,\\
g_{b0} &=& g_{b1}=f_{b1}=0\,.
\eeq
in very good agreement with respect to fully relativistic expansion from Table~\ref{tab:funcs} (notice that the fully relativistic expansion is restricted to only mildly Newtonian oscillatons).

For small $r$, $V_{1}$ is larger in magnitude than the Newtonian potential. This seemingly odd result was recognized in Ref.~\cite{UrenaLopez:2002gx}. The physical origin of this property can be traced to the scalar pressure, which is of the same order of magnitude as the energy density. We are thus in a weak-field but Newtonian-like limit. The gradient of the Newtonian potential is dominated by the magnitude of the gradient of the time-dependent potential for $\lambda r \leqsim  0.57Z$ and becomes an order of magnitude larger at $\lambda r \approx 1.06Z$~\cite{weakly}.

\section{Motion in dynamical spacetimes} \label{sec:motionformalism}

\subsection{Geodesics in the full geometry} \label{sec:motionfullgeo}
The action of a pointlike particle in the spacetime metric described by \eqref{eq:metric} is
\be
S[\gamma(\tau)]= \int_{\gamma(\tau)}L(x,\dot{x})= \int_{\tau_i}^{\tau_f}g_{\alpha\beta}\dot{x}^\alpha \dot{x}^\beta d\tau  \label{Lagrangean}\,,
\ee
where $\gamma(\tau)$ is a curve on the spacetime. The geodesics on this metric are the curves that extremize the action above, when varied with respect to the curve  $\gamma(\tau)$,
\begin{equation}
\ddot{x}^\mu+\Gamma^{\mu}_{\hphantom{\mu}\beta\alpha}\dot{x}^\beta\dot{x}^\alpha=0\,,
\end{equation}
where the Christoffel symbols are defined by
\begin{equation}
\Gamma^{\mu}_{\hphantom{\mu}\beta\alpha} \equiv \frac{g^{\mu \nu}}{2}\left(\partial_{\beta}g_{\nu\alpha} + \partial_{\alpha}g_{\nu\beta} - \partial_{\nu}g_{\alpha\beta}\right)\, .
\end{equation}
Since $\tau$ is an affine parameter we can always re-parametrize it such that $L=-1,0,1$ for
timelike, null or spacelike geodesics, respectively.

For the Lagrangean~\eqref{Lagrangean},  $\varphi$ is a cyclic coordinate. Thus, its conjugate momentum, the angular momentum along the $z-$axis $r^2\sin(\vartheta)\dot{\varphi}=J$, is a conserved quantity . Due to the spherical symmetry of the metric, the geodesics will always be planar. Therefore, without loss of generality, we set $\vartheta = \frac{\pi}{2}$. The geodesic equations are  reduced to two nontrivial coupled equations,
\beq
&&\ddot{t}+\frac{1}{2A}\left(\del_t A\dot{t}^2+ 2A'\dot{r}\dot{t}+\del_t B\dot{r}^2\right)=0 \,,\label{geodesics1}\\
&&\ddot{r}+\frac{1}{2B}\left(B'\dot{r}^2+A'\dot{t}^2- 2 r\dot{\varphi}^2 + 2\del_t B\dot{r}\dot{t}\right)=0 \,. \label{geodesics2}
\eeq

Two examples of trajectories concern circular and radial motion, for which 
\beq
r(\tau)&=& r_0\,,\, \dot{r}=0\,,\, \ddot{r}=0\\
r(\tau=0)&=&r_{\rm init}\,,\,\dot{\varphi}=0\,,
\eeq
respectively.

Consider first circular motion in our chart representation of the manifold. The substitution $\dot{\varphi} = \Omega$ in equation \eqref{geodesics2} yields,
\be
\frac {A'}{2B}\dot{t}^2-\frac{r_0\Omega^2}{B}=0\,.
\label{eq:dott0}
\ee
There is a solution  if $A'>0$. Solving Eq.~\eqref{eq:dott0} for $\dot{t}$ and differentiating to find $\ddot{t}$ we may rewrite equation~\eqref{geodesics1} as
\begin{equation}
\frac{\del_{t} A'}{A'}-\frac{\del_t A}{A}=0.
\end{equation}
Any non-null separable function $A(t,r)=a_t(t)a_r(r)$ satisfies this condition, making it sufficient for circular motion to be allowed.

For timelike ($L=-1$) motion we find,
\be
\Omega=\frac{1}{\sqrt{\frac{2r_0a_r(r_0)}{a^{'}_r(r_0)}-r^{2}_0}},\label{Angular_freq}
\ee
implying that $2a_r/a'_r>r_0$ at $r_0$. Note, incidentally, that the {\it coordinate} angular velocity is,
\be
\tilde{\Omega}=\frac{d\varphi}{dt}=\frac{d\varphi}{d\tau}\frac{d\tau}{dt}=\frac{\Omega}{\dot{t}}\,.
\label{eq:OmtotildeOm}
\ee
From \eqref{eq:dott0},
\be
\dot{t}=\Omega\sqrt{\frac{2r_0}{a'_r\left(r_0\right)a_t\left(r_0\right)}}\,,
\ee
thus we find
\be
\tilde{\Omega}=\sqrt{\frac{a'_r\left(r_0\right)a_t\left(r_0\right)}{2r_0}}\,, \label{eq:coordinatangfreq}
\ee
in agreement with standard results~\cite{Cardoso:2008bp}.

Following the same prescription, null geodesics must satisfy
\begin{equation}
r_0=\frac{2a_r(r_0)}{a^{'}_r(r_0)}\,.\label{constraintnull}
\end{equation}
When these conditions are applied to standard spacetimes, such as the Schwarzschild geometry, one recovers well-known results~\cite{Cardoso:2008bp}.

\subsection{Geodesics in weakly dynamic spacetimes}
Up to now, our results are generic and valid for any geometry of the form \eqref{eq:metric}. When the spacetime is {\it approximately}
static as in Eq.~\eqref{metric_expansion}, some simplifications occur.
Let $x^{\mu}_0(\tau)$ be the solution of the equations of motion derived from the Lagrangian 
$L_0(x^\mu,\dot{x}^\mu)=g^{(0)}_{\alpha\beta}\dot{x}^\alpha\dot{x}^\beta$. Assume $x^\mu_0(\tau)$ to be known.
Let us consider what happens to the solution when the metric is slightly perturbed as in Eq.~\ref{metric_expansion}, such that it gives rise to a new Lagrangian 
$L(x,\dot{x})=L_0(x^\mu,\dot{x}^\mu)+\epsilon L_1(x^\mu,\dot{x}^\mu)$. The associated geodesics will have a different solution, which should lie sufficiently close to $x^\mu_0(\tau)$. Lets find it, by expanding around $x_0^\mu$: 
\beq
&&x^\mu(\tau)=x_0^\mu(\tau)+\epsilon\,\eta^\mu(\tau)\,,\\
&&L(\eta,\dot{\eta},\tau)=L_0(x_0)+\epsilon L_1(x_0) + \epsilon \frac{\partial L_0}{\partial x^\alpha} \eta^\alpha+\epsilon \frac{\partial L_0}{\partial \dot{x}^\alpha} \dot{\eta}^\alpha\nonumber\\
&+&  \epsilon^2 \left(\frac{1}{2}\frac{\partial^2L_0}{\partial x^\alpha x^\beta}\eta^\alpha \eta^\beta+\frac{1}{2}\frac{\partial^2L_0}{\partial \dot{x}^\alpha \dot{x}^\beta}\dot{\eta}^\alpha \dot{\eta}^\beta+\frac{\partial^2L_0}{\partial {x}^\alpha \dot{x}^\beta}\eta^\alpha \dot{\eta}^\beta \right)\nonumber\\
&+& \epsilon^2 \frac{\partial L_1}{\partial x^\alpha} \eta^\alpha +\epsilon^2 \frac{\partial L_1}{\partial \dot{x}^\alpha}\eta^\alpha+{\cal O}(\epsilon^3)\,.
\label{expansion}
\eeq
Although not explicitly stated, each partial derivative of the Lagrangean is to be evaluated at $x_0(\tau)$, here and in the following.

Since $x_0(\tau)$ is known and solves the $0^{th}$ order equations of motion, the above expansion only depends on $\eta$, $\dot{\eta}$ and $\tau$. Thus, to extremize the action of this Lagrangean, the variation must be done in $\eta$. The Euler-Lagrange equations of ~\eqref{expansion} yield, up to second order in $\epsilon$, the following nontrivial result
%
\begin{equation}
\frac{d}{d \tau}\left(\frac{\partial \zeta}{\partial\dot{x}^i}  \right)-\frac{\partial \zeta}{\partial x^i}  = \frac{\partial L_1}{\partial x^i}  - \frac{d}{d\tau}\left(\frac{\partial L_1}{\partial \dot{x}^i} \right)\,,
\label{initgeoress}
\end{equation}
where we defined
\begin{equation}
\zeta =\frac{\partial L_0}{\partial x^j}\eta^j + \frac{\partial L_0}{\partial \dot{x}^j}\dot{\eta}^j\,.
\end{equation}
Expressing $L_0$ as a function of the metric, equation \eqref{initgeoress} can be further simplified into
%
\beq
&&\ddot{\eta}^\gamma+2\Gamma^{\gamma}_{(0)\alpha\beta}\dot{x}_0^\alpha\dot{\eta}^\beta+\left(\partial_\delta \Gamma^{\gamma}_{(0)\alpha\beta}\right)\dot{x}_0^\alpha\dot{x}_0^\beta \eta^\delta\nonumber\\
&&=-\frac{1}{2}g^{(0)\, \gamma \beta}\left(\frac{d}{d\tau}\left(\frac{\partial L_1}{\partial \dot{x}^\beta} \right) -\frac{\partial L_1}{\partial x^\beta} \right)\,.
\label{georessonance}
\eeq
%

If the LHS was equated to 0, one would have the geodesic deviation equation, describing the evolution of a perturbation on the geodesic itself. However, there is a ``force-term'' presented on the RHS illustrating how the metric's perturbations disturb the geodesics of the background static spacetime. This result can also be obtained by firstly applying Euler-Lagrange equations to $L(x,\dot{x})$ and only then considering the expansion around $x_0$.

By inspection of equation \eqref{initgeoress}, if $x_0^\gamma(\tau)$ is cyclic in $L_0$, the equation may be integrated on both sides, leading to a first order equation
%
\beq
\frac{\partial \zeta}{\partial\dot{x}^\gamma} &=& \frac{\partial^2L_0}{\partial \dot{x}^\gamma \dot{x}^\beta} \dot{\eta}^\beta + \frac{\partial^2L_0}{\partial \dot{x}^\gamma x^\beta} \eta^\beta\nonumber\\
&=&2g^{(0)}_{\gamma\beta}\dot{\eta}^\beta+2\partial_\beta (g^{(0)}_{\gamma\alpha})\dot{x}_0^\alpha\eta^\beta\nonumber\\
&=&\int_{\tau_0}^{\tau}\left(-\frac{d}{dy}\left(\frac{\partial L_1}{\partial \dot{x}^\gamma} \right) +\frac{\partial L_1}{\partial x^\gamma} \right)dy+C_\gamma,
\label{cyclicgeoress}
\eeq
%
where $C_\gamma$ is a real constant depending on the initial conditions.

\subsection{Motion in weakly-dynamic, time-periodic spacetimes} \label{sec:periodic_spacetimes}

We will now specialize the discussion to weakly-dynamic \textit{and} time-periodic geometries, like the one described by \eqref{eq:weakly1} and \eqref{eq:weakly2}.
For the background metric, both $t$ and $\varphi$ are cyclic coordinates in $L_0$, allowing the corresponding two equations in system \eqref{georessonance} to be rewritten as first order equations using \eqref{cyclicgeoress}:
\beq
\dot{\eta}^\varphi&=&-\frac{2}{r_0}\dot{\varphi}_0\eta^r+\frac{C_\varphi}{2r_0^2} +\frac{F_{\varphi}}{2r_0^2}\,,\label{cyclivarsphi}\\
\dot{\eta}^t&=&-\frac{a_0'}{a_0}\dot{t}_0\eta^r-\frac{C_t}{2a_0}-\frac{F_t}{2a_0}\,,
\label{cyclivarst}
\eeq
where
\beq
EL_\gamma &=& \left(-\frac{d}{d\tau}\left(\frac{\partial L_1}{\partial \dot{x}^\gamma} \right) +\frac{\partial L_1}{\partial x^\gamma} \right)\,,\\
{\cal F}_t&=&\int_{\tau_0}^{\tau} EL_t(y) \, dy\,,\quad {\cal F}_\varphi=\int_{\tau_0}^{\tau} EL_\varphi(y) \, dy\,,
\eeq
and $a_0(r_0) = g^{(0)}_{tt}(r_0)$, $b_0(r_0) = g^{(0)}_{rr}(r_0)$.
As we discussed previously, geodesics on the background metric $g^{(0)}$ defined in Eq.~\eqref{metric_expansion} are planar, thus allowing the choice $\vartheta_0(\tau)=\pi/2$. 
Replacing the relations \eqref{cyclivarsphi} and \eqref{cyclivarst} on the system \eqref{georessonance}, we are left with a system of two decoupled second-order equations for $\eta^r$ and $\eta^\vartheta$:
%
\beq
&&2 r_0   \left(\ddot{\eta}^\vartheta+\eta^\vartheta \dot{\varphi}_0^2\right)+4\dot{\eta}^\vartheta  \dot{r}_0=\frac{EL_\vartheta }{r_0 }\label{2ordertheta}\,, \\
&& \eta^r  \left(\dot{r}_{0}^2 b_0''+\dot{t}_{0}^2 a_0''+6 \dot{\varphi}^2\right)+2 \dot{\eta}^r  \dot{r}_0  b_0'+2 b_0 \ddot{\eta}^r\nonumber\\
&&- \frac{\eta^r  b_0'\left(\dot{r}_0^2 b'_0+\dot{t}_0^2 a_0'-2 r_0  \dot{\varphi}^2\right)}{b_0} = \frac{\dot{t}_0  a_0' \left({\cal F}_t+C_t+2 \eta^r  \dot{t}_0  a_0'\right)}{a_0}\nonumber\\
&&+\frac{2 \dot{\varphi}  \left({\cal F}_\varphi+C_\varphi\right)}{r_0}+EL_r  \,\label{2orderr}\,.
\eeq
%

\subsection{Symmetries of motion} \label{sec:symmetries}

Since the metric coefficients of a dynamic spacetime are time dependent, time homogeneity -- valid in the  time-independent spacetime -- no longer holds generically and time is not a cyclic coordinate. Specifically, in the case of time-periodic spacetimes, symmetry is not completely lost but reduced to a discrete subgroup, akin to (space) translation symmetry in crystals~\cite{crystals}.

Breaking of the time homogeneity has interesting consequences for the study of the test particle initial value problem; take for example the case of circular background motion. In the weakly-dynamic spacetime, the time-dependent part of the metric is treated as a perturbation. Assume therefore a metric expansion of the form \eqref{eq:weakly1} and \eqref{eq:weakly2}, where the background metric 
$g^{(0)}_{\alpha\beta}$ is static and spherically symmetric. If we want our ``initial time'' $t_{i}$ to correspond to a vanishing perturbation, we should set it to $\pi/(4\omega)$ or to $3\pi/(4\omega)$. The solution to the problem depends on the specific choice one makes, because of the different signs of the cosine derivatives in the time-dependent part of the metric. 
Note that this situation is consistent with the spherical symmetry of the problem: for initial times different from zero, the relation $\varphi \propto t$ is not valid anymore, instead $\varphi \propto (t-t_{i})$. In other words, it is irrelevant at what point on the initial orbit (at fixed times) our particle is when the perturbation is turned on; however, it is important at what point in time that happened. Time-periodic perturbations break the time homogeneity and the same initial conditions, but different $t_i$, don't lead to the solutions which can be related by the time translation in $t_i$.

In the quantum treatment of the electron motion in crystals, Bloch's theorem implies that there is a conservation of the crystal momentum. However, as the symmetry is discrete, Noether theorem is not applicable and the conservation law is a consequence of the linearity of Quantum Mechanics (Schrödinger equation is subject of the Floquet theory for the symmetry in question). As General Relativity is highly non-linear, we should not expect that point particle energy will be periodic in $\pi/\omega$, in analogy with electrons in crystals, in general. We can calculate the change of the particle energy function $E(t)=-\partial L/\partial \dot{t}$, at the order ${\cal O}(\epsilon^2)$, between two arbitrary moments in time
\beq 
\label{eq:energy_periodicity_general}
&& E(t_2)-E(t_1)= \nonumber\\
&& 2\epsilon \omega  \int^{\tau(t_2)}_{\tau(t_1)}\left(b_1(r_{0})\dot{r}_{0}^2-a_1(r_{0})\dot{t}_{0}^2\right)\sin{(2\omega t_0(y))} dy\,.
\eeq
The integrand is not necessarily periodic in $\pi/\omega$, so we can't claim $E(t)=E(t+n\pi/\omega),n \in \mathbb{N}$. However, this conclusion will change when the equations of motion become linear, as in Section \ref{sec:circularmotionexample}.

We should also note that spacetimes with metric expansion as in the example do admit time-inversion symmetry. This symmetry is broken when friction is present, as in Section~\ref{sec:dissipation}.

 \section{The excitation of resonances}  \label{sec:examples}

\subsection{Circular and radial background motion - linear regime \label{sec:circularmotionexample}} 
%

We start by considering the perturbations presented in \eqref{eq:weakly1} and \eqref{eq:weakly2} on background circular geodesics. Imposing the circularity condition ($\dot{r}_0=0, \ddot{r}_0=0$) in Eq.~\eqref{geodesics2} one finds
\be
r_0(\tau)=r_0\,,\quad   \varphi_0(\tau)=\Omega\tau\,,\quad t_0(\tau)=\frac{\Omega}{\tilde{\Omega}}\\
\tau+t_i\,,\label{Circular Solution}
\ee
where $\tilde{\Omega}$ is given by equation \eqref{eq:coordinatangfreq} and $t_i$ is a constant. The requirement that the motion is timelike is equivalent to requiring that $\Omega$ be given by Eq.~\eqref{Angular_freq}. Furthermore, $EL_\vartheta$ and $EL_\varphi$ vanish; replacing $EL_t$ and $EL_r$ by their corresponding expressions, an analytical solution is found for $\eta^r(\tau)$  through \eqref{2orderr}:
\be
\eta^r(\tau) = \mathcal{D}(\omega) \cos \left(2 \omega t_0(\tau)\right)+C_1 \cos (\Theta \tau + C_2) + C_3\,.
\label{circularsol}
\ee
where $C_1$ and $C_2$ are real constants dependent on the initial conditions of $\eta^r$, $C_3$ is related with the initial conditions of $\eta^t$ and $\eta^\varphi$, and:
\beq
\mathcal{D}(\omega) &=& \frac{r_0 \left(a_0 a_1'-a_1 a_0'\right) }{8 r_0 \omega^2 b_0 a_0+2 r_0 (a'_0)^2-a_0 \left(r_0 a_0''+3 a_0'\right)}\,, \nonumber\\
\Theta &=& \Omega  \sqrt{\frac{r_0 a_0 a_0''-2 r_0 a_0'^2+3 a_0 a_0'}{b_0a_0a_0'}}\,.\label{Theta}
\eeq
It is clear that there may exist {\it resonances} in the motion, when the amplitude $\mathcal{D}(\omega)$ diverges. This occurs at frequencies
$\omega=\omega_{\rm res}$ for which the denominator of $\mathcal{D}(\omega)$ above vanishes. We find
{\begin{equation} \label{eq:circresonant}
\omega_{\rm res}=\pm\frac{\Theta}{2\dot{t}_0}\,.
\end{equation}}

The frequency $\Theta$ corresponds to the proper radial epicyclic frequency for this static, axially symmetric spacetime \cite{RonaldoVieira:2017}. 
To obtain the radial epicyclic frequency in coordinate time we need to divide $\Theta$ by $\dot{t}_0$. Note that the frequency of the metric perturbation in \eqref{eq:weakly1} and \eqref{eq:weakly2} is in fact $2\omega$. Then, the effective frequency of resonance corresponds to $2\omega_{res}=\Omega/\dot{t}_0$ which, as stated, is the radial epicyclic frequency in coordinate time.
Thus, our system behaves as a classic, driven harmonic oscillator: when the ``forcing'' frequency equals the natural (epicyclic) frequency, a resonance occurs. 
If $\Theta$ differs from $\Omega$, the geodesics will precess. The above is a very generic prediction of a smoking-gun of time-periodic spacetimes.

Finally, let us apply \eqref{eq:energy_periodicity_general} to this specific background motion. As $r_0$ and $\dot{t}_0$ do not depend on the proper time, the integral reduces to zero when $t \rightarrow t+n\pi/\omega,n \in \mathbb{N}$. This conclusion is not valid during the resonant motion when the higher order terms become important.

We now  focus on perturbations on radial geodesics. Imposing $\dot \varphi_0(\tau)=0$ and $\ddot \varphi_0(\tau)=0$, an explicit analytic solution for $\eta^r$ is not possible for general radial geodesics. Thus we specialize to motion of small amplitude around the geometric center of our spacetime. Expanding $L_0$ to first order around $r=0$ and $\dot{r}=0$, the geodesics following from $L_0(x,\dot x)$ admit the following solution:

\beq 
r_0(\tau)&=&\tilde{r}_0\cos\left(\Omega_0 \tau\right)+\frac{\dot{ \tilde{r}}_0}{\Omega_0}\sin\left(\Omega_0\tau\right)\,,\\
t_0(\tau)&=&\alpha\tau+t_i\,,\\
\Omega_0&=&\alpha\sqrt{\frac{a_0''}{2b_0}}\,,\quad\alpha=\dot{t}_0(\tau)\,,\label{Radial Solution}
\eeq
where $\tilde{r}_0$ and $\dot{\tilde{r}}_0$ are initial conditions. Although not explicitly stated, all quantities are to be evaluated at $r=0$. In the preceding derivation, we used the fact that the parity of $a_0$, $b_0$, $a_1$ and $b_1$ implies that these functions and their odd radial derivatives vanish at the origin, for regular spacetimes.

Expanding \eqref{2orderr} on $r$ and $\dot{r}$ around 0, using the appropriate expressions for $EL_\varphi$, $EL_t$ and $EL_r$, we obtain the following analytic solution,
\beq
\eta^r(\tau)&=&-\mathcal{Q}(\omega)\frac{\dot{r}_0(\tau ) \cos (2 \omega \alpha  \tau  ) 
}{(\alpha  \omega ) \left(4 b_0 a_0\left(2 b_0\omega ^2-a_0''\right)\right)}\nonumber\\
&-& \mathcal{G}(\omega)\frac{r_0(\tau) \sin (2 \omega \alpha  \tau   ) 
}{4 b_0a_0 \left(2 b_0 \omega^2-a_0''\right)}\nonumber\\
&+&C_1 \cos (\omega_0 \tau  +C_2)-C_t\frac{\dot{r}_0(\tau )}{2 \alpha  a_0} \tau\,, \label{nrradial}
\eeq
$C_1$ and $C_2$ are constants depending on the initial conditions on $\eta^r$. $C_t$ is the integration constant in \eqref{cyclivarst} and
\beq
\mathcal{G}(\omega)&=&b_0\left(a_0a_1''-a_1a_0''\right)+b_1a_0 a_0''\,,\\
\mathcal{Q}(\omega)&=&4 b_0b_1 a_0\omega^2+\mathcal{G}(\omega)\,.
\eeq
The solution grows linearly in time unless $C_t=0$.  This is an expected result as, by inspection of equation \eqref{cyclivarst}, one may conclude that, for a non vanishing $C_t$, $\dot{\eta}^t$ is given by a sum of a constant with the integral of a periodic function, inducing a linear growth of $\eta^t$. As all our equations depend on a small and stable evolution of the motion, this result may be alarming. Nevertheless one may always choose $C_t$ to be null, as it is only related to the initial conditions of $\dot{\eta}^t$ which are decoupled of the initial perturbation on the radial direction. For these reasons, we will use $C_t=0$. 
 
Again, there is a frequency $\omega$ for which the two denominators on \eqref{nrradial} vanish, corresponding to a resonance. This frequency is
\begin{equation}\label{eq:freqrad}
\omega_{\rm res}=\sqrt{\frac{a_0''(0)}{2b_0(0)}}=\frac{\Omega_0}{\alpha}\,.
\end{equation}
Now, resonance occurs when the perturbation frequency $(2\omega)$ is two times the "natural" frequency by which $r_0(\tau)$ oscillates. This result is extremely intuitive, if one imagines the spacetime pushing the object away from the centre, while the latter is also going away from it, and pulling inwards when the object starts moving towards the centre. Then, in each half period of the small oscillation in $r_0$, the metric's perturbation must complete a full period. Similarly to the circular case, \eqref{eq:energy_periodicity_general} implies energy periodicity in $\pi/\omega$ for the radial motion, when we concentrate on small amplitude deviations ${\cal O}(\tilde{r}_0^2)$.

\subsection{Non-relativistic motion in the weak-field regime} \label{sec:N_dynamics}

In order to precisely establish the analogy with the driven harmonic oscillator from the previous section, and to understand dynamical aspects beyond the linear regime, we consider non-relativistic particle motion in the weak field limit. The Lagrangian for the test particle non-relativistic equatorial motion in a weak and asymptotically flat spherically-symmetric spacetime, like the one given by \eqref{eq:NexpA} and \eqref{eq:NexpB} is described by
\begin{equation} 
2L=-(1+\nu)\dot{t}^2+\dot{r}^2+r^2\dot{\varphi}^2.
\end{equation}
Here $\nu(t,r)=2V(r)+\epsilon 2V_{1}(r)\cos(2 \omega t)$. $V(r)$ is the Newtonian gravitational potential and $V_1(r)$ originates from time-dependent part of the $A(t,r)$ metric coefficient. 
These quantities are defined in Eq.~\eqref{eq:NexpA2} and are related to the original metric coefficients as $1+2V=a_0$ and $2V_1=a_1$.

The Euler-Lagrange equations reduce to:
\begin{align} 
  & -(1+\nu)\ddot{t}=\frac{1}{2} \partial_t \nu\dot{t}^2+ \nu' \dot{t}\dot{r},\\
  & \partial^2_t r+\partial_t r\frac{\ddot{t}}{\dot{t}^2}=-\frac{1}{2} \nu'+r\tilde{\Omega}^2. \label{eq:EL_weak_field_radial_0}
\end{align}
As we are interested in the non-relativistic motion $\dot{r}<<\dot{t}$
\begin{align} 
  & \partial^2_t r-\frac{\tilde{J}^2}{r^3}=-\frac{1}{2}\nu', \label{eq:ELweakfieldradial}
\end{align} 
where we have introduced the coordinate angular momentum $\tilde{J}=r^2\tilde{\Omega}$. We see that the second term on the l.h.s. of \eqref{eq:EL_weak_field_radial_0} is of order $\sim v/c^2$, when we restore $c$. These equations are valid even for the highly-dynamical spacetime, as we will further elaborate in the Section~\ref{sec:Mathieu_rapid}.

Now we focus on the weakly-dynamical spacetime $(\epsilon<<1)$  and weak orbital perturbations from background circular motion $r=r_0+\epsilon \eta^r  $ as in \eqref{Circular Solution}. Equation \eqref{eq:ELweakfieldradial} then reduces to:
\begin{align} 
& \tilde{\Omega}=\sqrt{\frac{1}{r_0}V'}, \\
& \partial^2_t\eta^r+(V'' +3\tilde{\Omega}^2 )\eta^r=-V_1'\cos{(2\omega t)} \label{eq:ELweakfieldcircpert}. 
\end{align}
In the last equation and until the end of this and the next section both potentials and their derivatives, with respect to $r$,  are  to  be  evaluated  at $r_0$. Last equation represents equation of motion for the driven linear harmonic oscillator, as claimed. Resonance occurs when
\begin{equation} \label{eq:N_res_freq}
\omega_{\rm res}=\frac{1}{2}\sqrt{V''+3 \tilde{\Omega}^2 }\,.
\end{equation}
This result agree with the appropriate limit of Eq.~\eqref{Theta}.

For radial motion $(\tilde{J}=0)$, equation \eqref{eq:ELweakfieldradial} yields:
\begin{align} 
& \partial^2_t r_0=-V', \label{eq:EL_weak_field_radial_pert_0th} \\
& \partial^2_t\eta^r+V'' \eta^r=-V_1'\cos{(2\omega t)} \label{eq:EL_weak_field_radial_pert_1st}. 
\end{align}
The solution of these equations depends on the form of the potential. However (see also Section~\ref{sec:circularmotionexample}), we can expand around the initial state $\{ r_0=0, \partial_t r_0=0 \}$ and for small amplitudes $\tilde{r}_0$ obtain
\begin{equation} 
r_0(t) = \tilde{r}_0\cos\left(\tilde{\Omega}_0 t \right)+\frac{\dot{\tilde{r}}_0}{\tilde{\Omega}_0}\sin\left(\tilde{\Omega}_0 t \right)\,,
\end{equation}
with $\tilde{\Omega}_0=\sqrt{V''(0)}$. Expanding $V_1(0)$ to the first non-zero term (as $V'(0)=V_1'(0)=0$, see Section~\ref{sec:circularmotionexample}), \eqref{eq:EL_weak_field_radial_pert_1st} becomes
\begin{equation} 
\partial^2_t\eta^r+\tilde{\Omega}^2_0 \eta^r=-V_1''(0)\cos{(2\omega t)r_0(t)}\,.\label{eqeta2}
\end{equation}
When we solve this equation, we see that resonance occurs when
\begin{equation} \label{eq:N_res_freq_radial}
\omega_{\rm res}=\tilde{\Omega}_0\,.
\end{equation}
This result coincides with the relativistic one~\eqref{eq:freqrad}.

\subsection{Higher order corrections, instabilities and the Mathieu equation} \label{sec:Mathieu}

The motion described by \eqref{eq:ELweakfieldradial} is formally the same as that of a point particle (in Newtonian gravity) around a spherical body whose luminosity changes~\cite{Saslaw:1978}~\footnote{Such scenario is relevant for the analysis of dynamics of dust or small planetary systems' bodies around variable stars, where the time-dependent radiation pressure, acting as a perturbation, influences orbital dynamics.}. In the following, we take a similar approach in order to assess the dynamics beyond the linear regime.

In the previous sections, we used a linear expansion 
$x^{\mu}=x^{\mu}_0+\epsilon \eta^{\mu} $ in the small parameter $\epsilon$ to understand the evolution of the perturbation. This expansion is in fact a truncated version
of the correct full series $x^{\mu}=x^{\mu}_0+\sum^{\infty}_{n=1}\eta^{\mu}_{n} \epsilon^n$. To understand what new features can arise in the full theory,
we now expand in the radial coordinate, still at the linear level, but with different parameter strength $\lambda < \epsilon$ - This will allow us to ``effectively'' probe the higher-order terms. Using this expansion, the equation of motion for the $\eta^r$ [Eqs.~\eqref{eq:EL_weak_field_radial_pert_1st} and \eqref{eq:ELweakfieldcircpert}] now  reduces to
\beq 
&& \lambda \partial^2_t\eta^r+\lambda \Big((2\omega_{\rm res})^2+\epsilon V_1 '' \cos{(2\omega t)} \Big)\eta^r \nonumber\\
&=& -\epsilon F\cos{(2\omega t)} +\mathcal{O}(\lambda^2)\,, \label{eq:ELweakfieldcircpertMathieu}
\eeq
where $F=V_1 '$ for circular orbits and $F=V_1''r_0(t)$ for radial motion is the forcing term. The corresponding resonance frequencies $\omega_{\rm res}$ are given by Eqs.~\eqref{eq:N_res_freq} and \eqref{eq:N_res_freq_radial} for circular and radial motion respectively. For $\lambda=\epsilon$, we recover the previous results at linear order in these parameters. The $\epsilon\lambda$ term in \eqref{eq:ELweakfieldcircpertMathieu} impacts the equations of motion of $\eta^{r}_2$, thus explaining our claim that we are ``probing'' higher order behaviour. From now on, we absorb $\epsilon$ in $V_1$ and $\lambda$ in $\eta^r$. Equation \eqref{eq:ELweakfieldcircpertMathieu} is known as the inhomogeneous Mathieu equation.

We can classify the motion described by \eqref{eq:ELweakfieldcircpertMathieu}, by comparing the driving and natural frequencies $\omega$ and $\omega_{\rm res}$ respectively. The motion can then be in {\it adiabatic} $(\omega_{\rm res} \gg \omega)$, nearly {\it parametric-resonant} $(\omega_{\rm res} \sim \omega)$ or {\it rapidly oscillating background} $(\omega_{\rm res} \ll \omega)$ regime. In the next few subsections we will focus on the circular background motion, but the analysis is easily generalized.

\subsubsection{Adiabatic regime} \label{sec:Mathieu_adiabatic}

Here it is natural to use an adiabatic approximation \cite{LLbookmechanics, binney2011galactic} in order to understand the motion of particles. For time scales of the order of $1/\omega_{\rm res}$, Eq.~\eqref{eq:ELweakfieldcircpertMathieu} describes an harmonic oscillator with a constant driving force. This type of external force does not deform the phase portrait of the oscillator, but only shifts it by a constant amount $\eta^r_{c,0}=-V_1 '/W(0)^2$ (think of the mass attached to the vertical elastic spring), where  
\be
W(t)=\sqrt{(2\omega_{\rm res})^2+V_1 '' \cos{(2\omega t)}}\,.
\ee
The time dependence will, on the one hand, modify the center of the phase-space trajectory [non-homogeneous term in \eqref{eq:ELweakfieldcircpertMathieu}] as $\eta^r_c \approx (W(0)/W)^2\eta^r_{c,0}\cos{(2\omega t)}$. On the other hand, the phase-space trajectory will be itself deformed because of the time dependence of the effective frequency $W$ [homogeneous part of \eqref{eq:ELweakfieldcircpertMathieu}]. The Hamiltonian that effectively describes the $\{\eta^r,\partial_t\eta^r \}$ motion is
\begin{equation} 
H=\frac{1}{2}(\partial_t\eta^r)^2+\frac{1}{2}W^2(\eta^r-\eta^r_{c})^2=IW\,,\label{eq:hamiltonian_adiabatic}
\end{equation}
where we introduced the action-angle variables $\{ I,\theta \}$ as $\eta^r-\eta^r_c=\sqrt{2I/W} \sin{\theta}$, $\partial_t\eta^r=\sqrt{2IW} \sin{\theta}$. As the action is approximately preserved during the adiabatic process we can calculate it at the initial time $I(0)=I_0$ and find approximate analytical solution to the \eqref{eq:ELweakfieldcircpertMathieu} in this regime
\begin{equation} \label{eq:adiabatic_solution}
\eta^r(t) \approx \eta^r_c(t)+\sqrt{\frac{2I_0}{W}} \sin{\theta(t)},
\end{equation}
where $\theta(t) \approx  (2\omega_{\rm res})t$ and we used Hamiltonian equations of motion $\partial_t \theta \equiv \partial H / \partial I= W$. From \eqref{eq:hamiltonian_adiabatic} we can also see the leading behavior of the energy function - it will be periodic with the period of $\pi/\omega$.


\subsubsection{Parametric resonances} \label{sec:Mathieu_parametric}
When the driving and natural frequencies are similar, parametric resonance can occur. These resonances happen when $W(t)=2\omega$. The dominant resonance is $\omega=\omega_{\rm res}$, but others can have important cumulative effects in the energy transfer and can make the orbit unstable~\cite{Saslaw:1978}. The stability of the homogeneous Mathieu equation is a well studied problem (e.g. Ref.~\cite{benderbook}). The inhomogeneity of the Mathieu's equation does not change the conclusions of this stability analysis if the driving term is harmonic with the same driving frequency as the one in $W$~\cite{CairncrossPelster:2014}. We will introduce dimensionless time and rescale parameters in \eqref{eq:ELweakfieldcircpertMathieu} as  $t = 2 \omega t$, $a=(\omega_{\rm res}/\omega)^2$, $2\epsilon=V_1 ''/(2 \omega)^2$ and $f=-V_1 '/(2 \omega)^2$:
\begin{equation} \label{eq:inhomogen_Mathie}
\partial^2_t\eta^r+(a+2\epsilon \cos{t})\eta^r=f\cos{t}.
\end{equation}
The stability of Mathieu's equation can be represented on the parametric $\epsilon-a$ (Ince-Strutt) stability diagram  (see Fig. 11.11 in Ref.~\cite{benderbook}). For most values of the parameters $a$, there are regions of $\epsilon$ where the solution to the Mathieu's equation are bounded and the corresponding orbits of particles are stable. However, if $a=n^2/4$ $(n \in \mathbb{N})$, instability occurs for any $\epsilon \neq 0$ \cite{benderbook}. These values are in terms of the driving frequency terms expressed as 
\beq 
 \omega&=&2\omega_{\rm res}/n \nonumber\\
&=& \{ 2\omega_{\rm res},\omega_{\rm res}, 2\omega_{\rm res}/3, \omega_{\rm res}/2, 2 \omega_{\rm res}/5, \rm{...} \}. \label{eq:N_omega_instable}
\eeq
It should be noted that these results do not depend on the sign of $\epsilon$, as Ince-Strutt stability diagram is symmetric under reflections about $a$ axis. 


\subsubsection{Rapidly oscillating background}
\label{sec:Mathieu_rapid}

Motion in a rapidly oscillating background is effectively dictated by the static background, because the perturber acts so rapidly that the system doesn't have time to adapt, similarly to the sudden approximation in Quantum Mechanics. This is the case, as we shall see, even when the ``perturbing'' force is the same order of magnitude as the ``non-perturbing'' force. Because of this,  we will be general and start the discussion by rewriting  \eqref{eq:ELweakfieldradial} as
\begin{equation} \label{eq:NII_law_osc}
\partial^2_t r = - U'(r)+F(r)\cos{(2\omega t)},
\end{equation}
where $U'(r)=V'(r)-\tilde{J}^2/r^3$ and $F(r)=-V_1'(r)$. Let the radial coordinate be decomposed as $r=r_{s}+\xi^r$, where $r_{s}$ and $\xi^r$ are slowly and rapidly varying parts, respectively. After this decomposition, the slow and rapid parts of the equation of motion \eqref{eq:NII_law_osc} must be separately satisfied \cite{LLbookmechanics}\footnote{One should be careful with the initial conditions when there is a non-zero phase \cite{RidingerDavidson:2007}.}. Rapid part will have the form
\begin{equation}
\partial^2_t\xi^r=F(r_s)\cos{(2\omega t)}+\mathcal{O}(\xi^r),
\end{equation}
where we assumed that $\xi^r$ terms are small. This equation can be easily integrated
\begin{equation} \label{eq:osc_motion_eta}
\xi^r(t)=-\frac{1}{(2\omega)^2} F(r_s)\cos (2 \omega t)
\end{equation}
and we can see that $\xi^r$ is indeed small, because of $1/\omega^2$ suppression, and that our assumption that the coordinate can be perturbatively decomposed irrespective of the fact that the ``perturbative'' force is bigger than the ``non-perturbative'' is correct. 
The slowly varying part of Eq.~\eqref{eq:NII_law_osc}, after averaging, has the form
\begin{equation}
\partial^2_t r_{s}=-U'(r_s)-\frac{1}{2(2\omega)^2} F(r_s)F'(r_s),
\end{equation}
As claimed, motion is governed by the time-independent effective potential and time-varying part is suppressed by $1/\omega^2$.

\subsection{Numerical evolution in the homogeneous background} \label{sec:toy_model}
%
\begin{figure*}[th]
\begin{tabular}{cc}
\includegraphics[width=0.45\textwidth,clip]{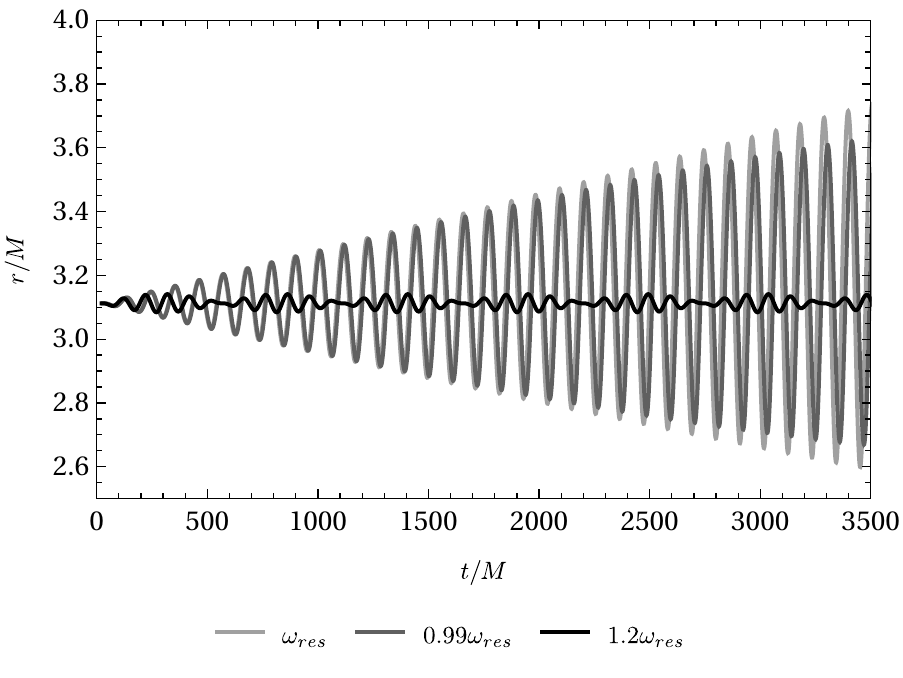}&
\includegraphics[width=0.45\textwidth,clip]{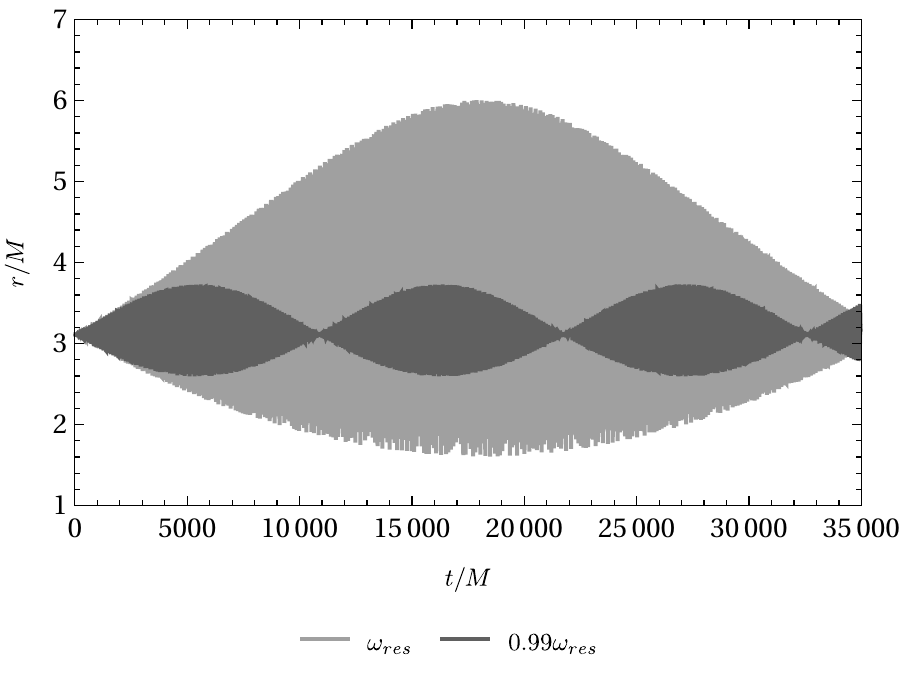}
\end{tabular}
\caption{Evolution of an initially circular geodesic in the spacetime of a time-periodic geometry~\eqref{eq:toyA}-\eqref{eq:toyB}, for different spacetime frequency $\omega$. The geodesic was circular in the static geometry of a constant density star with radius ${\mathcal C}=0.1$, placed at an initial radius $r=3.11 M$. Because the full geometry is now time-dependent with $\epsilon=10^{-3}$, the motion is not perfectly circular nor closed. For this example, there is a resonance at $M\omega=M\omega_{\rm res}=34.7\times 10^{-3}$. It is apparent that as $\omega$ is tuned in close to resonance the motion differs wildly from its unperturbed circular trajectory.}
	\label{fig:GeoNumCirc}
\end{figure*}
\begin{figure*}[th]
\centering
\includegraphics[scale=1.3]{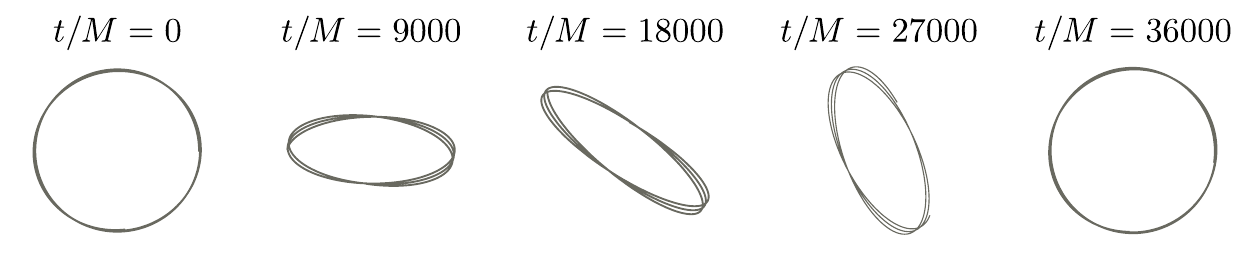}
\caption{Parametric representation of the geodesic corresponding to resonance; the central object is the same as that in Fig.~\ref{fig:GeoNumCirc}. The values of $t/M$ are the initial instant for which the geodesic is being represented, during $\delta t/M=500$.  As stated, the motion is not perfectly circular nor closed. The geodesic oscillates periodically from the initial orbit to an ellipse of large eccentricity, returning to the unperturbed motion after a beating period.
}
\label{fig:ParametricCirc}
\end{figure*}

To explore these results in a specific setup, we solved the unperturbed equations of motion numerically in an artificial, toy-model:
a constant density star spacetime, on top of which a time-periodic fluctuation was superposed. To be specific, we set the metric components:
\beq
A&=&A_{\rm star}(r)+\frac{M^2\,A_{\rm star}(r)}{M^2+r^2}\epsilon \cos(2\omega t)\,, \label{eq:toyA} \\
B&=&B_{\rm star}(r)+\frac{M^2\,B_{\rm star}(r)}{M^2+r^2}\epsilon \cos(2\omega t)\,,\label{eq:toyB}
\eeq
Here, $A_{\rm star}(r), B_{\rm star}(r)$ correspond to the geometry of a constant density star, of mass $M$ and radius $R$ in General Relativity~\cite{bookWeinberg:1972, Macedo:2013jja}. This rather arbitrary choice could mimic for example radially oscillating stars or other geometries. For us here, it is merely a toy arena where we can test of previous results.
We assume that there is no coupling between the fluid in the star and the orbiting object, and that therefore the object follows a geodesic. A straightforward analysis shows that
for $\epsilon=0$ there are {\it stable} circular timelike geodesics for any $r^2<R^3/(2M)$ and that they are all stable if the compactness $\mathcal{C}<23/54$.

In the weak field limit of our toy model \eqref{eq:toyA}, $A_{\rm star}(r)=1+2V_{\rm star}(r)$ and
\begin{equation} \label{eq:N_star_potential}
V_{\rm star}=-\mathcal{C}\Big(\frac{3}{2}- \frac{r^2}{2R^2}\Big)
\end{equation}
corresponds to the potential of a homogeneous sphere in the Newtonian gravity (or spherical harmonic oscillator). From \eqref{eq:N_res_freq} we obtain $\omega_{\rm res}=\tilde{\Omega}$. This result is consistent with the evaluation of \eqref{eq:circresonant} for dilute relativistic stars described by \eqref{eq:toyA}, when $\Theta \approx 2 \Omega$. Note that in this setup the homogeneous part of \eqref{eq:ELweakfieldcircpert} is the result expected from Newtonian gravity - precession occurs for the radial perturbations of circular motion inside the homogeneous sphere with the epicyclic frequency $2\tilde{\Omega}$ \cite{binney2011galactic}.

For the relativistic numerical investigation we took a star with compactness $\mathcal{C}=0.1$ and $\epsilon=10^{-3}$. Using the above, the functions $a_0\,,\,a_1\,,\,b_0\,,\,b_1$ [defined in \eqref{eq:weakly1}-\eqref{eq:weakly2}] are trivially known, and the geodesics can be numerically solved without approximations, using the full metric. We imposed initial conditions corresponding to fully unperturbed circular geodesics,
and monitored the position $r(t)/M$. The trajectory is shown in Fig.~\ref{fig:GeoNumCirc} for three different ``driving'' frequencies $\omega$. 
Since the equations of motion are accurate up to order ${\cal O}(\epsilon^2)$, an absolute resonance is not featured in our fully numerical solution. Instead, we find a beating pattern due to the interference of the two sinusoidal signals in \eqref{circularsol}. We applied a numerical Fourier analysis to these solutions to understand the spectrum of frequencies present in the data.
Our results show a clear, discrete spectrum of two frequencies for each example. These match, to within an error of $~0.1\%$, to the ones given by $\Theta(\omega)$ in \eqref{Theta} and $\omega$,  confirming the validity of our perturbative analytic results. The beating frequency is defined by the half difference between the frequencies of the two signals in \eqref{circularsol}
\beq \label{eq_beating_freq}
\omega_{\rm beat}=\frac{\omega-\omega_{\rm res}}{2}\,.
\eeq
Therefore, the beating period becomes larger as $\omega$ approaches $\omega_{\rm res}$ , given by Eq. \eqref{eq:circresonant}, as seen in Fig.~\ref{fig:GeoNumCirc}. Likewise, the position $r(t)$ grows to larger 
amplitudes as $\omega$ reaches $\omega_{\rm res}$, indicating that this is, in fact, a resonance.

To understand whether instabilities, as predicted by the analysis of 
Sections \ref{sec:Mathieu_parametric}, were possible, we numerically evolved the orbits for ``driving'' frequencies $\omega$ given by \eqref{eq:N_omega_instable} and explored the parameter space spanned by $\{ \mathcal{C},\,\epsilon,\,r_0,\, t_i\}$. The motion is confined to within the ``star'' at all times, i.e. $r(t)<R$. For all the parameter values that we explored, we found {\it resonant-like} behaviour for $\omega=2 \omega_{\rm res}$. For $\omega= \omega_{\rm res}/2$, the behaviour depends on the
value of the parameters. When $\mathcal{C} \sim 0.1$, small $r_0$ and large $\epsilon$ we found resonant-like behavior. However, for other points in the parameter space, the envelope of the $r(t)$ is seemingly linearly and very slowly growing. This growth may be tamed at some proper time, but we haven't observed this in all cases. Regarding the phase $t_i$, the existence of resonances seems
to be independent on it.

We have not found any resonant or unbounded motion for $\omega=2\omega_{\rm res}/3$, $\omega=2\omega_{\rm res}/5$ or $\omega=\omega_{\rm res}/3$. As this conclusions haven't changed for dilute backgrounds (i.e. weak fields) where $\mathcal{C} \sim 10^{-3}$ we can conclude that higher order terms of the expansion are responsible for the taming of the  instabilities predicted by the Mathieu equation. It should be noted that we numerically evolved trajectories until $\tau/M \sim 10^{9}$ and that resonant or unbounded motion may become apparent at later times in the cases where it was not found. 
The orbital motion in the adiabatic and rapidly-oscillating background regime is in very good agreement with the behaviour described by Eq.~\eqref{eq:adiabatic_solution} and in Section \ref{sec:Mathieu_rapid}, respectively, for small and qualitatively even for large compactness~\footnote{In these two regimes, the equations of motion are stiff: they contain two time scales with big gaps between them and one should use appropriate integrators \cite{PressRecipes:2007}.}.
\begin{figure}
\centering
\includegraphics[width=\columnwidth]{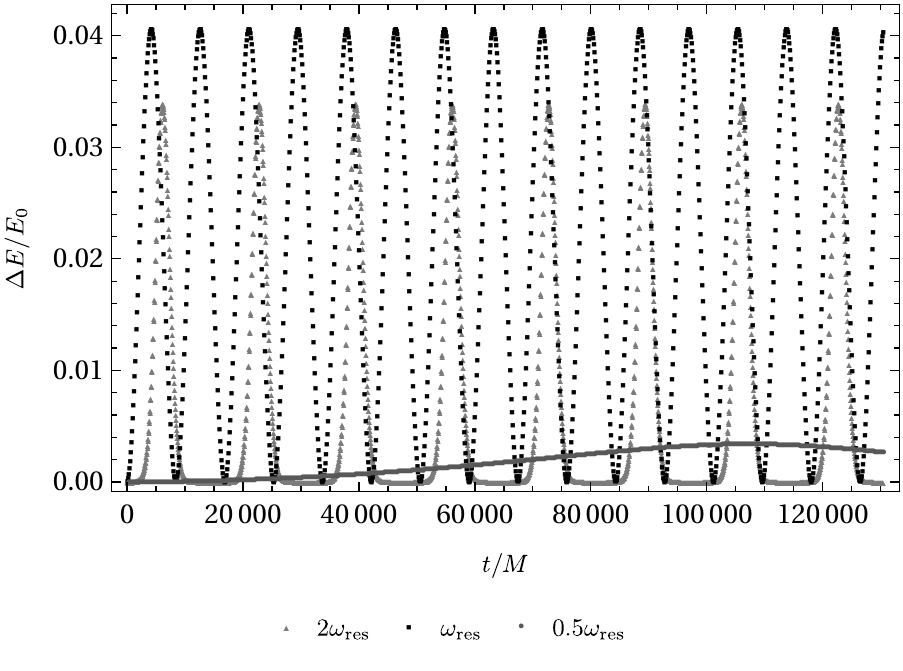}
\caption{Time evolution of the variation of the orbit's energy ($E=-\partial L/\partial t$), weighted by its initial value. The particle has a background circular motion around the toy model under the same conditions that were simulated in fig. \ref{fig:GeoNumCirc}, with $\epsilon = 1/100$. The perturbative term has frequencies, $2\omega_{\rm res}$, $\omega_{\rm res}$ and $.5\omega_{\rm res}$, for which the evolution of the energy is periodic with the frequency of the envelope on resonant motion.}
\label{fig:EnergyvTime_circular}
\end{figure}

The time evolution of the energy, $E=-\partial L/\partial t$, is as expected from our earlier general considerations. In particular, focus on the behaviour of $\Delta E=|E(n\pi/\omega)-E_0|$, $n \in \mathbb{N}$ and $E_0=E(0)$. When the driving frequency is not given by $\omega=\{\omega_{\rm res}/2,\omega_{\rm res},2\omega_{\rm res} \}$, we find $\Delta E/E_0 \ll 1$. On the other hand, when $\omega$ equals one of these resonant frequencies, the relative change is periodic, as seen in Fig.~\ref{fig:EnergyvTime_circular}. We find that the period is the same as the envelope of $r(t)$ in Fig.~\ref{fig:GeoNumCirc}.

Finally, the parametric representation of the geodesic in Cartesian coordinates, shown in Fig.~\ref{fig:ParametricCirc}, features the predicted precession of the geodesic. The initial circular geodesic is deformed into an ellipse whose eccentricity peaks when the deviation is also maximum, before returning to circular after a beating period.

We have also studied radial motion in this background. The features of the motion strongly depend on the parameters. In general, if the time-varying component of the metric is strong enough
(i.e. for large enough $\epsilon$), resonances can be excited for {\it any} initial conditions {\it and} for any background frequency. We observed such behavior for sufficiently dilute configurations.
The natural frequency of small-amplitude oscillations in general differs from the background's. However, the spacetime drives the object to frequencies
which seem to be a multiple of background's.
This drift in frequency is confirmed by numerical Fourier analysis and the phenomena is observed both for ``driving'' frequencies much larger and smaller than the natural frequency (we find a drift even when the driving frequency is two order of magnitude larger or smaller). The behaviour of the solution in this scenario departs strongly from the one described in Section~\ref{sec:circularmotionexample}; however, all the resonances are tamed at some point in proper time. This behavior seems to be strongly model-dependent. In the small $\epsilon$ regime, our results
are similar to the circular case.

\subsection{Motion in oscillatons} \label{sec:applications_oscillatons}

We now briefly discuss how the previous results apply to oscillaton spacetimes, studied in Section~\ref{sec:Oscillatons}.
The motion in the spacetime describing oscillatons has been studied numerically for very specific conditions~\cite{2006GReGr..38..633B}, the results obtained agree qualitatively with the conclusions drawn from Section~\ref{sec:toy_model} for homogeneous backgrounds, and with our own simulations of motion in oscillatons. In Ref.~\cite{2006GReGr..38..633B}, background circular geodesics were perturbed and numerically solved for an oscillaton, with initial conditions corresponding to a turning point. For different values of initial radius and angular momentum, bound orbits with elliptic-like behavior were observed, similar to the ones in our toy model example of Fig.~\ref{fig:ParametricCirc}. Both the period and amplitude of the oscillation were sensible to these conditions, as seen in both our analytical and numerical results. Ref.~\cite{2006GReGr..38..633B} also found that there are initial conditions for which the amplitude of oscillations are negligible.
Our results agree with these findings.

An obvious question concerns the existence of resonances for these objects. We used our numerical results from Section~\ref{sec:ROscillatons} to compute the ratio between the resonant and the oscillaton's frequency, for both circular, Eq.~\eqref{eq:circresonant}, and radial, Eq.~\eqref{eq:freqrad} motion. We paid also special attention to multiples of such ratio, for which the Mathieu equation predicts instabilities - Eq. \eqref{eq:N_omega_instable}. Such ratio, by virtue of being dimensionless, can only depend on the product of oscillaton $M$ and the scalar $\mu$ mass, and the compactness $\mathcal{C}$ is a suitable choice of dimensionless combination. For the most compact oscillatons $(\mathcal{C} \sim 0.07)$, the difference between $\omega$ and $2\omega_{res}$ was a factor of two too large, and the gap widens as $\mathcal{C}$ decreases.
In summary, our results indicate that neither on circular nor radial motion is able to excite resonances in oscillaton spacetimes.

We can use the analytical solution in the Newtonian regime (Section~\ref{sec:NOscillatons}) to confirm these findings for dilute oscillatons. Taking the expansion~\eqref{eq:Nmetricexpansion} and using it in Eq.~\eqref{eq:freqrad}, we find that, for radial motion near the origin the resonance frequency is
\beq
\omega_{\rm res}\left(\mathcal{C}\right)=1.33791 \, \mathcal{C}\,.
\eeq
Therefore, the resonance frequencies are bounded from above by the maximum allowed compactness. For the maximum compactness for which the Newtonian analysis is valid, $\mathcal{C}\approx 0.01$, one finds $\omega_{\rm max}=0.013379$. This upper bound is considerably {\it lower} than the oscillatons's frequency of Fig.~\ref{fig:OvsPhi0} (even in the relativistic scenario).
Thus no resonances are excited by radial motion.

For circular motion, using Eq.~\eqref{eq:circresonant} we find,
\beq
\omega_{\rm res}(\mathcal{C})=\frac{\mathcal{C}}{2\sqrt{2}}\sqrt{\frac{14.32 - 88.6408 \, \mathcal{C}}{1 + \mathcal{C} \,(-6.19 + 1.79\, r^2  \, \mathcal{C})}}\,.
\eeq
Thus, we find again an upper bound
\beq
\omega_{\rm res}<\frac{\mathcal{C}}{2\sqrt{2}}\sqrt{\frac{14.32}{1 -6.19\, \mathcal{C}}}\,.\label{eq:freqcircmaj}
\eeq
The r.h.s grows with compactness between 0$<\mathcal{C}<0.161551$. This means that for the Newtonian regime ($\mathcal{C}<0.01$) the frequency of resonance in circular background motion is bounded by the r.h.s of \eqref{eq:freqcircmaj} evaluated at $\mathcal{C}=0.01$, which is $\omega_{\rm max}= 0.0138134$. Once again, this value does not get near the frequencies of oscillatons corresponding to the Newtonian regime in Figure \ref{fig:OvsPhi0}. We conclude that the motion in Newtonian oscillatons is in the rapidly oscillating background regime $(\omega \gg \omega_{\text{res}})$. It is, thus, appropriate to use the formalism developed in Section \ref{sec:Mathieu_rapid}, to understand motion in these oscillatons, applicable irrespective of the fact that spacetime is highly-dynamical or not.

We can focus the discussion on homogeneous oscillatons where, via \eqref{eq:NPoisson}, \eqref{eq:NV2}  and \eqref{eq:N_star_potential},
\beq
V'(r)=-\frac{1}{3}V_{1}'(r)=\frac{M}{R^3}r.
\eeq
This description is, by \eqref{eq:Ndensitysmallr}, valid for motion near the center. The amplitude of the rapidly varying part of the radial coordinate  \eqref{eq:osc_motion_eta} is proportional to $\tilde{\Omega}^2/ \omega^2$, where $\tilde{\Omega}^{-1}=\sqrt{R^3/M}$ is the dynamical time scale of the slowly-varying radial component. If the object has $\mathcal{C}=10^{-2}$, on the verge of the weak-field limit validity, then (see also Fig. \ref{fig:res_self_int_osc}) $\tilde{\Omega}^2/\omega^2 \sim 10^{-4}$. Thus, the motion is always well described by a smooth transition between background (circular or radial) motion on which small perturbations are superposed.

We also studied motion in the presence of scalar self-interaction, at the Newtonian level, but we find the same qualitative behavior (see Appendix \ref{AppGPP_self_int_motion}). The attractive self-interaction makes the configuration more compact in the inner core, with respect to the non-self-interacting one. For highly compact oscillatons the gap between $2\omega_{\rm res}$ and $\omega$ is not too large, and the Newtonian analysis raises the intersting issue of whether or not such self-interaction term could lead to resonances in highly-compact oscillatons. 
Such question is specially interesting for the full axion potential~\cite{Helfer:2017a}.

\subsection{The effect of dissipation on resonances} \label{sec:dissipation}
%
\begin{figure}[th]
\includegraphics[width=0.5\textwidth]{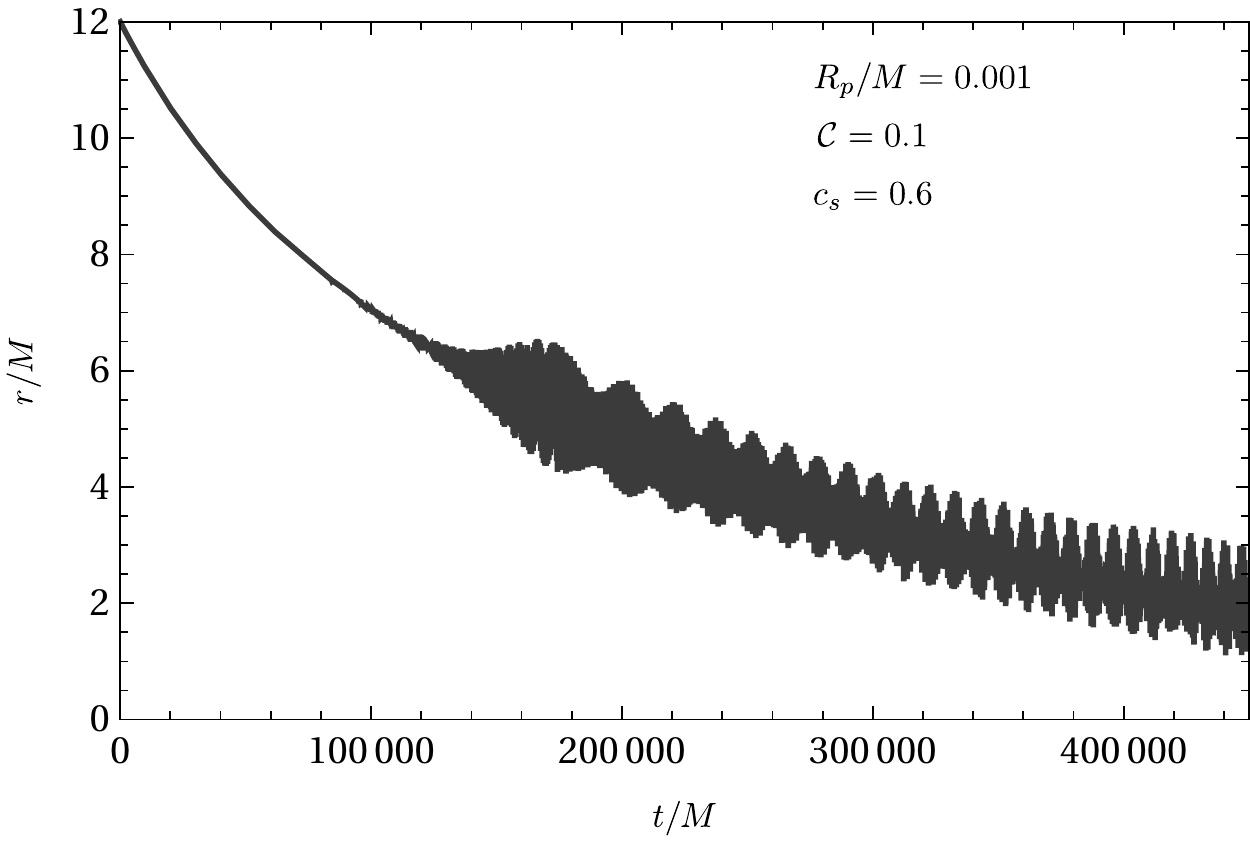}
\caption{Radial motion of an object subject to very weak dissipation on the time-periodic geometry \eqref{eq:toyA}-\eqref{eq:toyB}, where resonance occurs for $r/M = 6$. The star has $\mathcal{C}=0.1$, the inspiralling object has $R_p = 10\, \mu_{p_0} =10^{-3}M$. The sound speed was taken to be $c_s=0.6 c$.\label{fig:DragGeoNumCirc}}
\end{figure}
The above analysis neglects dissipative effects. It is, in principle, possible that the resonances do not leave any observable imprint
in realistic situations: gravitational drag, along with gravitational radiation losses could, for instance, drive the inspiralling body inwards without
even being affected by resonances. To test this, we have added a dissipative force $F$ to the motion of the body, of initial mass $\mu_p(\tau=0)=0$ and radius $R_p$. The equations of motion are given by
\beq
\mu_p\left(\ddot{x}^{\gamma} + \Gamma^{\gamma}_{\alpha\beta} \dot{x}^{\alpha} \dot{x}^{\beta}\right) = F^{\gamma}.
\label{eq:dragmotion}
\eeq 
The force can describe several effects, such as gravitational radiation reaction, accretion of gravitational drag~\cite{Macedo:2013qea,Barausse:2014tra}. Regarding accretion, for a small compact object (radius $R_p$ much smaller than the mean free path) its mass growth is determined by
\beq
\dot{\mu}_p = \frac{\pi \rho R_p^2}{v},
 \label{eq:accretion}
\eeq 
where $\rho$ is the density of the compact object ``generating'' the dynamical spacetime configuration and $v$ the relative velocity between the orbiting body and the compact star. The gravitational drag force may be modelled by dynamical friction on a constant-density medium~\cite{Macedo:2013qea,Barausse:2014tra}
\beq
F_{DF}=- \frac{4 \pi \mu_p ^2 \rho}{v^2} I_v\,,
\eeq
with, 
\beq
I_v = 
\begin{cases}
\frac{1}{2}\log \left(\frac{1 + v/c_s}{1 - v/ c_s}\right) - v/c_ s , \quad &v < c_s \\
\frac{1}{2}\log \left(1-\frac{c_s^2}{v^2}\right) + \log \left(\frac{v t}{r_{\rm min}}\right) , \quad &v > c_s \\
\end{cases}
\eeq
where $c_s$ is the velocity of sound in this medium and $r_{\rm min}\sim R_p/v$~\cite{Macedo:2013qea}.
In our simulations, we always used subsonic motion.

Choosing once again $\vartheta = \pi/2$ and taking into account the effect of accretion and gravitational drag, the dissipation force may be modelled in a Newtonian way:
\beq
F_D^t &=& 0\,,\\
F_D^r &=& -\dot{\mu}_p\dot{r} + F_{DF} \frac{\dot{r}}{v}\,,\\
F_D^\varphi &=& -r\dot{\mu}_p\dot{\varphi} + F_{DF} \frac{r\dot{\varphi}}{v}\,.\label{eq:dissipationequation}
\eeq
The system \eqref{eq:dragmotion} and \eqref{eq:accretion} determines the motion of an object through this perturbed spacetime.

Fig.~\ref{fig:DragGeoNumCirc} represents the radial evolution of an originally circular motion of a very small object in the previous homogeneous toy model, undergoing subsonic dissipation. As in the Section \ref{sec:toy_model}, the compactness of the star is taken to be $\mathcal{C}=0.1$, the density of the medium $M^2\rho=3M^3/(4 \pi R^3)\simeq2.38732 \cdot 10 ^{-4}$ and the speed of sound ($c_s$) was chosen to be $0.6 c$. Regarding the orbiting object, its initial mass and radius was chosen to be very small on the geometry ($R_p = 10 \,\mu_{p_0} = 10^{-3}M$), such that the effect of the drag would be small. The oscillating frequency of this toy model was chosen such that resonance would occur at $r/M = 6$. It is clear from Fig.~\ref{fig:DragGeoNumCirc} that the object undergoes a very slow inspiralling motion until it reaches $r/M \sim 6$. Then, the eccentricity of the orbit lowers, and the trajectory becomes similar to the resonant behaviour studied in the previous section. This implies that an object captured by the gravity of this periodic structures will undergo a resonance (if possible) when reaching the correspondent radius, enabling its observation with appropriate techniques. 

Our results for larger damping indicate that the drag hastens the decay of the object to the center of the star. For large enough damping, the resonance (and forcing) has little impact on the motion, as expected. Consequently, if the friction is too large, resonance might be unobservable.

\section{Implications for ultra-light axion DM halos} \label{sec:darkhalo}

We will now use the previous results to investigate whether the motion in the background of the ultra light axion DM can lead to observable consequences.
Previous works have considered the impact of FDM on motion of binary pulsars~\cite{Khmelnitsky:2013lxt, Blas:2016ddr, DeMartinoBroadhurst:2017} and laser interferometers~\cite{AokiSoda:2017a}. We are here interested not in the impact of FDM on the motion of binary pulsars around their center of mass but, for illustration, motion of that center of mass around the galactic halo. We discuss the former aspect of motion in Appendix \ref{AppBinary}.
Masses of axion particles in the range $10^{-23}\leqsim m[\text{eV}] \leqsim 10^{-27}$ are of interest in the context of mixed axion DM (MDM)~\cite{HlozekMarsh:2017}.

\subsection{Halo description} \label{sec:darkhalo_descr}
Galactic DM halos in the FDM scenario are stabilized by Heisenberg's uncertainty relation (or quantum pressure in a hydrodynamical perspective, see Appendix \ref{AppGPP}) and are related to the weak-field limit of the EKG system, studied in Section~\ref{sec:NOscillatons} and Appendix \ref{AppEKGNewt}. However, the connection between Newtonian oscillatons and DM halos is not straightforward. It is theoretically expected that a dark halo consists of a nearly homogeneous core surrounded by particles which are behaving like CDM \cite{MarshPop:2015}. Simulations of galaxy formation with FDM confirm such picture~\cite{SchiveChiueh:2014, Mocz:2017}. The density profile of such effectively cold, DM region is described by the standard Navarro-Frenk-White (NFW) profile~\cite{binney2011galactic}. In the case of the FDM, this core
is usually referred to as a pseudo-soliton~\footnote{``Pseudo'' because it is not protected by a charge. However, it is stable on cosmological scales (see Section \ref{sec:ROscillatons}).}. Simulations reveal non-local scaling relations between the parameters that describe the soliton and the whole halo \cite{Schive:2014hza, Mocz:2017, 4caveat}.

There are established procedures for constructing FDM halos and reconstructing observational galactic rotation curves~\cite{MarshPop:2015,GonzalezMoralesMarsh:2017, Bernaletal:2017a}. The dark halo is described by
\begin{equation}  \label{eq:halodensity}
\rho(r)=\rho_{\text{sol}}(r)\theta(r_{\epsilon}-r)+\rho_{\text{NFW}}(r)\theta(r-r_{\epsilon})\,,
\end{equation}
where
\begin{equation} \label{eq:solitondensity}
\rho_{\text{sol}}(r)=\frac{\rho_c}{(1 + 0.091(r/r_c)^2)^8} \,,
\end{equation}
and
\begin{equation}
\rho_{\text{NFW}}(r)=\frac{\rho_s}{\frac{r}{r_s}\big(1 + \frac{r}{r_s} \big)^2}\,,
\end{equation}
are soliton and NFW profiles, respectively, and $\theta$ is the Heaviside function. Here, $\rho_c$ is the central density of the soliton, $r_c$ (the core radius) is the point at which the density falls off to half of its central value, $\rho_s$ is related to the density of the Universe at the moment the halo collapsed, $r_s$ is NFW scale radius, $r_{\epsilon}$ is soliton-NFW transition radius. Demanding continuity of the soliton and NFW densities at the transition (and optionally their first derivative), we are left with only four (three) free parameters which can be found by fitting galactic rotation curves. 

The soliton density function~\eqref{eq:solitondensity} was found by fitting onto results of galaxy formation simulations~\cite{SchiveChiueh:2014}. The fitted density distribution \eqref{eq:solitondensity} for the soliton is in excellent agreement with our approximate analytical solution of Section~\ref{sec:NOscillatons}. One of the two soliton parameters $(\rho_c, r_c)$ can be replaced instead by the axion particle mass $m$~\footnote{We remind the reader that $m=\mu\hbar$.}. This is a global parameter independent of the galactic details. From the definition of $r_c$ and the scaling in Eq. \eqref{eq:NSPscale},
\begin{equation} \label{eq:fuzzy_cdensity}
\rho_c=1.94 \times 10^{-2} \Big ( \frac{r_c}{1\text{kpc}} \Big )^{-4} \Big ( \frac{m}{m_{22}} \Big )^{-2}  \frac{M_{\odot}}{\text{pc}^3}.
\end{equation}
In the last equation $m_{22}=10^{-22} \text{eV}$ and we used our analytical profile to obtain the numerical prefactor. Simulations indicate that the transition radius usually corresponds to $r_{\epsilon} \approx 3.5 r_c$ \cite{Mocz:2017}. The scale-invariant radius of that point is $Z_s \equiv \lambda \mu r_{\epsilon} \approx 1.035 Z$. In order to make a connection with our description of oscillatons in Section \ref{sec:Oscillatons}, we also find how their compactness\footnote{Here, we define $\mathcal{C_{\text{s}}}=M_s/r_{\epsilon}$, where $M_s=M(r_{\epsilon})$. We are not considering whole Newtonian oscillatons, because in FDM halos the exponential tail is substituted with the NFW tail. From the scaling relations $\mathcal{C}=\mathcal{C_\text{s}}(Z_s/Z)(\beta/\beta_s)$ and using the full expansion we find $\beta_s=1.725$.} depends on the core radius and the particle's mass
\begin{equation}
\label{eq:compactness_halo} 
\mathcal{C_{\text{s}}} = 3.08 \times 10^{-9} \Big ( \frac{r_c}{1\text{kpc}} \Big )^{-2} \Big ( \frac{m}{m_{22}} \Big )^{-2}.
\end{equation}
Note that rotation curve fits indicate changes of  $r_{c}$ by at most two orders of magnitude $r_c \sim (10^{-2},1) \text{kpc}$ for FDM particle's masses interval \cite{Bernaletal:2017a}. If we use parameters for the Milky Way (MW), estimated in Refs.~\cite{SchiveChiueh:2014,DeMartinoBroadhurst:2017}: $m=0.8m_{22}$ and $r_c=120 \text{pc}$, we find $\rho_c=146 M_{\odot}/\text{pc}^3$ and $\mathcal{C_{\text{s}}}=3.34 \times 10^{-7}$. 

\subsection{Halo spacetime dynamics and general features of motion} \label{sec:darkhalo_motion}

In Section \ref{sec:NOscillatons} we showed that the spacetime describing the soliton is highly dynamical, in the sense that the gradient of the time-dependent potential (time-dependent force) $|V'_{1}|$ is of the same order of magnitude as or larger than the Newtonian gravitational force $V'$. Regarding the halo ``atmosphere'' (outer layer), as $r_s \sim r_\epsilon$ \cite{Bernaletal:2017a}, we are interested in the $r\gg r_s$ limit of the NFW profile $\rho_{\text{NFW}}(r) \sim \rho_s ( r/r_s )^{-3}$. In this limit, the time-dependent force becomes {\it logarithmically} smaller than the Newtonian force $V'/|V'_{1}| \sim \ln(r/r_s)$. The dark halo radius, taken as $r_{200}$, the point when the mean halo density is $200$ times bigger than the cosmological critical density, is usually two orders of magnitude larger than $r_s$. Thus, even at the halo radius the dynamical component is of the same order of magnitude as the static component. We can conclude that the whole halo is highly dynamical.

As we saw in Section~\ref{sec:applications_oscillatons}, the motion in solitons is in a rapidly oscillating background regime, as $\tilde{\Omega} \ll \omega$, where $1/\tilde{\Omega}$ is timescale associated with the motion dictated by Newtonian force. The ratio of the celestial object's Keplerian orbital and the oscillaton frequency depends on the core radius and the particle's mass near the soliton center\footnote{A useful number to keep in mind is that the period of oscillation for ULA with $m=m_{22}$ is $T \sim 1 \text{yr}$.} [from Eq. \eqref{eq:fuzzy_cdensity}]
\begin{equation} \label{eq:fuzzy_omega_ratio}
\frac{\tilde{\Omega}}{\omega} \approx 4 \times 10^{-9}  \Big ( \frac{r_c}{1\text{kpc}} \Big )^{-2} \Big ( \frac{m}{m_{22}} \Big )^{-2}\,.
\end{equation}
For our reference MW parameters, we find $\tilde{\Omega}/\omega \sim 4.34 \times 10^{-7}$. As rotation curves in outer regions of the halo ``atmosphere'' develop plateaus, the orbital frequencies must go further down at large distances. The effect of the time-dependent force on the orbital motion in this regime is, as explained in Sections~\ref{sec:Mathieu_rapid} and \ref{sec:applications_oscillatons}, suppressed by $(\tilde{\Omega}/\omega)^2$. In the absence of other matter sources, such suppression is extremely large in the galactic context (for the above mentioned estimates for MW, this is equal to $10^{-13}$) irrespective of the highly dynamical nature of the spacetime.

This order-of-magnitude reasoning neglects the presence of other matter (baryons and CDM) components. Let us assume rigid body rotation and estimate on which scales resonances -- in the presence of such other matter components -- may occur
\begin{equation} \label{eq:fuzzy_circluar_velocity}
r \sim 2 \times 10^{-6}   \Big ( \frac{v}{10\text{km}/\text{s}} \Big )  \Big ( \frac{m}{m_{22}} \Big )^{-1} \text{pc}\,.
\end{equation}
%
Thus, for typical orbital velocities resonances are not encountered on $\text{kpc}$ scales for neither the FDM nor the MDM mass range. We need to rely on the gravity of SMBH, and we address this in the next section.


As we saw, the impact of the time-dependent force on the object's position is suppressed, but ``only'' by $\tilde{\Omega}/\omega$ [as one can see from Eq.\eqref{eq:osc_motion_eta}]. Can one test this effect in astrophysics? To understand this issue, we now estimate the amplitude of the radial velocity oscillation with respect to the value dictated by the time-independent forces, with the Doppler effect in mind. This ratio is 
\beq \label{eq:Doppler_ratio}
&&\frac{\Delta v}{v}(r)=2.78\times 10^{-7}\Big ( \frac{m}{m_{22}} \Big )^{-1}\frac{\rho_{\text{ULA}}(r)}{10^2\frac{M_{\odot}}{\text{pc}^3}}  \\ \nonumber
&& \Big(\frac{r}{10^2\text{pc}}\Big )^{\frac{3}{2}} \Big(\frac{M_\text{tot}(r)}{10^{9} M_\odot}\Big)^{-\frac{1}{2}},
\eeq
where $M_\text{tot}(r)$ corresponds to the total (DM+baryon) enclosed halo mass\footnote{We model baryon component of the MW Galaxy as a bulge+disk. As we are interested in a simple model, we will describe bulge with Hernquist profile \cite{binney2011galactic}, having total mass of $\sim 1 \times 10^{10} M_{\odot}$ and a scale length $\sim  1 \text{kpc}$ \cite{WidrowDubinski:2005}. For the description of (thin) disk we used 3MN profile, where MN stands for Miyamoto-Nagai disk, with the parameters for the potential taken from Ref. \cite{RoryChris:2015} and a total mass of  $4.6 \times 10^{10} M_{\odot}$.}. 
For MW parameters given in Section~\ref{sec:darkhalo_descr}, the value of this ratio is $\sim 10^{-7}$ outside SMBH gravitational influence radius $\sim 2\text{pc}$ \cite{MerrittD:2010} and inside $r \leqsim 200 \text{pc} $. What happens for even lower axion masses, in the MDM domain? $M(r)$ is dominated by baryons in the first few $\text{kpc}$ and is unchanged. For a toy model of the dark halo in mixed DM cosmologies we take $\rho_{\text{DM}}(r)=\rho_{\text{ULA}}(r)+\rho_{\text{CDM}}(r)$, where ULA component $\rho_{\text{ULA}}(r)$ is described by \eqref{eq:halodensity} and CDM component $\rho_{c}(r)$ by a separate NFW profile~\cite{AnderhaldenDiemand:2013}. These two profiles are related by demanding that their relative abundance is equal to the cosmological one\footnote{Relative cosmological abundance of CDM and axions is given by $f_{i}=\Omega_{i}/\Omega_{d}$, where $i=a,c$ can stand for cold $\Omega_{c}$ and axion $\Omega_{a}$ components and total dark sector density is given by   $\Omega_{d}=\Omega_{a}+\Omega_{c}$. Present CMB constraint allow for $f_{a} 	\leq 0.53$ for $m=10^{-24} \text{eV}$,   $f_{a} 	\leq 0.33$ $(m=10^{-25} \text{eV})$, $f_{a} 	\leq 0.05$ $(m=10^{-26} \text{eV})$ and $f_{a} 	\leq 0.03$ $(m=10^{-27} \text{eV})$ \cite{HlozekMarsh:2017}.} 
at $\sim r_{200}$. For our simple estimate, the cosmological abundance ratio is taken to be the same throughout the halo, approximately~\cite{AnderhaldenDiemand:2013} and that the NFW parameters are the same for cold DM component as the one in the pure-CDM MW~\cite{SalucciNesti:2013}. With these assumptions in mind, for $m \sim 10^{-24} \text{eV}$, the ratio in Eq. \eqref{eq:Doppler_ratio} would have the value $\sim 10^{-6}$ for stars inside $r \leqsim 30 \text{pc} $. For orbital velocities of the order of $100 \text{km}/\text{s}$, this would require unrealistic helioseismology-level \cite{Brito:2015yfh} precision of $0.1\text{m}/\text{s}$ in order to be detected over $100\text{yr}$. For $m \sim 10^{-27} \text{eV}$ \eqref{eq:Doppler_ratio} can be as high as $\sim 10^{-4}$, however oscillation timescales become very large $\sim 10^5 \text{yr}$. Note that these radial velocities refer to the halo center and that the Solar System itself would experience radial velocity fluctuation, albeit much more suppressed at the Solar galactocentric distance of $8 \text{kpc}$.

\subsection{Constraining ULA density at the Galactic center} \label{sec:darkhalo_res}
We now show that the motion of bright S stars can be used to constraint ULA DM densities at the center of our galaxy. 
Data from the last 20 years was used to probe Yukawa-like modifications of gravity \cite{HeesPRL:2017}, and new data is expected to be of further help in this endeavour~\cite{HeesELT:2017}. This year the S0-2 star will be at its closest distance from the SMBH, and a redshift measurement is expected~\cite{HeesProc:2017}.

The behaviour of matter in the sub-parsec region is dominated by the SMBH gravity. We should stress that our understanding of DM behaviour in the presence of the SMBH and during the galactic evolution timescales is still in its infancy and mostly focused on CDM~\cite{ GenzelEisenhauer:2001, MerrittD:2010} (but see also Refs.~\cite{Hui:2016ltb,Ferreira:2017pth, Helfer:2017a, LopezAlbores:2018, 3caveat}). The core density estimate from Section~\ref{sec:darkhalo_descr} may not apply in the sub-parsec region, since the SMBH can make the region denser, e.g. by adiabatic growth \cite{MerrittD:2010}. Thus, we use constraints from previous analysis of the orbit of S stars as a rough upper limit on the extended background and treat fraction of ULA component as a free parameter. Present constraints allow for $1\sigma$ upper limit of $M_{\text{ext}}=10^{-2}  M_{\text{SMBH}}$, where $ M_{\text{SMBH}}=4.02 \times 10^6 M_{\odot}$ and background radius cutoff was fixed at $R=11 \text{mpc}$ in order to encompass whole of S0-2 star orbit~\cite{BoehleGhez:2016}. Some CDM estimates in this region are of order $\sim 10^3 M_\odot$ \cite{GhezSalim:2008}. This background consists of faint stars, compact objects and DM. We arbitrarily take the maximum contribution of ULA, $\lambda_{\text{ULA}}$, to be $30\%$. In this approach we don't have prior restrictions on the axion mass range that we probe but orbital timescales focus the range on FDM and MDM.

To estimate the impact of ULA time-dependent force, consider a simplified model where stellar orbits are influenced only by the (non-rotating) SMBH at the Post-Newtonian (PN) level and all other matter components are incorporated in the homogeneous spherically-symmetric extended background $\rho_{\text{ext}}$. The stellar equations of motion are, from Sections \ref{sec:NOscillatons} and \ref{sec:N_dynamics} and Refs.~\cite{RubilarEckart:2001, poissonwillbook}
\beq 
\partial^2_t \vec{r} &=&-\frac{1}{r^3}\Big[\Big(1+4\frac{1}{rc^2}+\frac{v^2}{c^2}\Big)\vec{r}-4\vec{v}\frac{\vec{v} \cdot\vec{r}}{c^2}\Big]\nonumber\\
&-&\frac{M_{\text{ext}}(r)}{r^3}\vec{r}+4\pi  \lambda_{\text{ULA}} \rho_{\text{ext}}\vec{r}\cos{(2\omega t+2\Upsilon)}\,.\label{eq_motion_smbh}
\eeq
In the above we use dimensionless quantities,
\be \label{eq_variables}
t =\frac{t}{\tau_{\text{dyn}}}\,,\quad r =\frac{r}{10\text{mpc}}\,,\quad M =\frac{M}{M_{\text{SMBH}}}\,,
\ee
$\tau_\text{dyn}=\sqrt{GM_{\text{SMBH}}/r_{I}^3}^{-1}=7.4\text{yr}$ is the dynamical timescale associated with the gravity of SMBH at $r_{I}=10\text{mpc}$ and $\Upsilon$ represents phase difference. 
In the context of our results from Sections \ref{sec:circularmotionexample}-\ref{sec:toy_model} for the circular example, the presence of the additional SMBH potential does not change the picture significantly, as it can be incorporated in $V(r_0)$. 

Most studied S stars around Sgr $\text{A}^{\star}$ are on a highly eccentric orbits, and the detailed treatment of such motion is outside the scope of this work. For such orbit, Eq. \eqref{eq:N_res_freq} is not applicable as the resonant frequency can change by as much as one order of magnitude along the orbit. We will onward focus only on S0-2 star. Its (initial) orbital parameters are $a_{0}=4.878(8)\text{mpc}$, $e_{0}=0.892(2)$ and $T_{0}=15.92(4)\text{yr}$ \cite{BoehleGhez:2016}. 
Present constraints on periastron precession are: $|\dot{w}_{0}|<1.7\times10^{-3}\,\text{rad}/\text{yr}$~\cite{HeesPRL:2017,HeesProc:2017}.
We have numerically solved the equations of motion, for different extended mass $M_{\text{ext}}$ and ULA abundance $\lambda_{\text{ULA}}$ as well as phase difference $\Upsilon$.
The system starts from the apocenter, and we monitor the secular (osculating) orbital parameters, averaging over one orbit. The motion is contained within a fixed orbital plane, even at the PN level \cite{poissonwillbook}, and, ipso facto, orbital inclination and longitude of the ascending node are fixed, which leaves us with semi-major axis, eccentricity, orbital period and periastron precession. From this set of orbital elements, only periastron precession is affected by PN effects \cite{poissonwillbook} and the homogeneous background \cite{RubilarEckart:2001, ZakharovNucita:2007, JiangLin:1985}, when a time-dependent ULA force is neglected. 

\begin{figure}
\centering
\includegraphics[width=1\columnwidth]{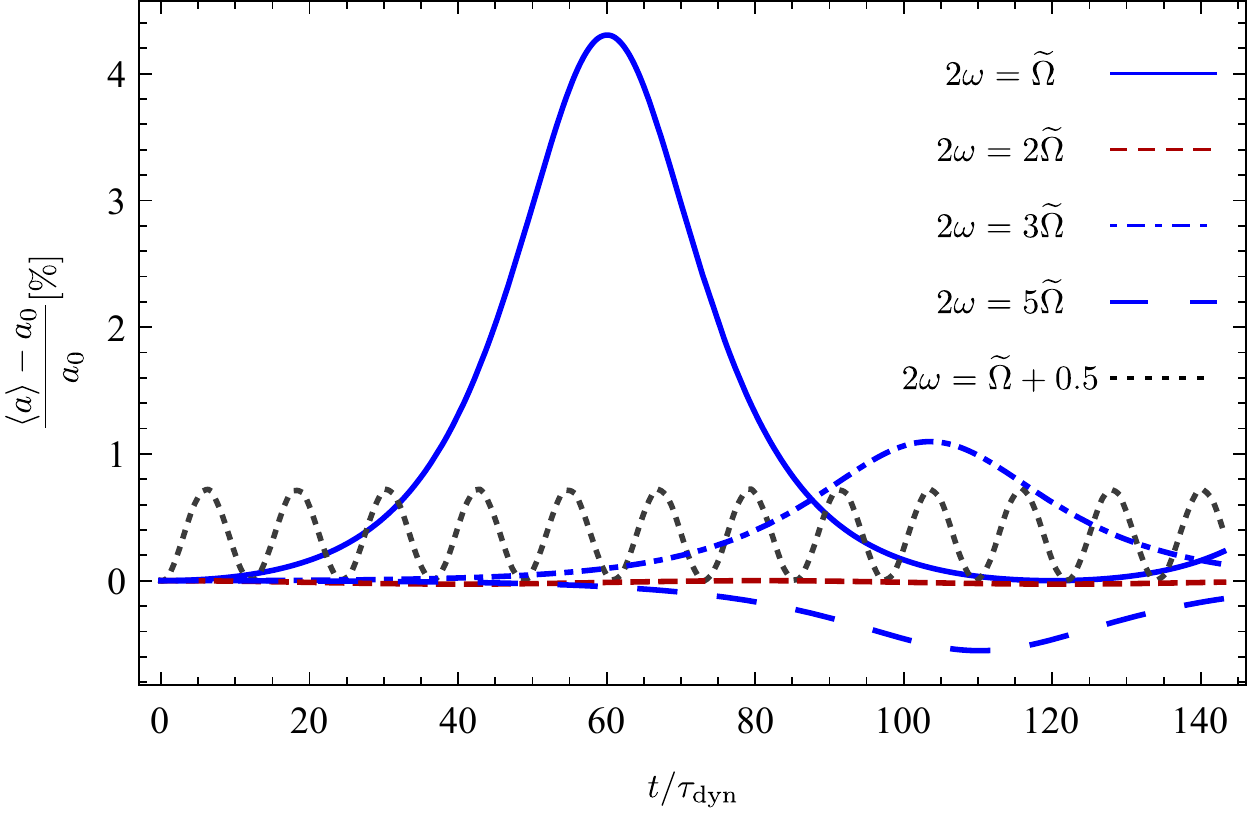}
\caption{Relative change of secular semi-major axis $a$ with respect to the value without present time-dependent force $a_0$ for three dominant resonant (blue; solid, dashed and dashed-dotted lines), one non-resonant (dashed red line) and one near-resonant (dashed black line) frequencies. The background is, for illustrative purposes, taken as the least conservative one: $M_{\text{ext}}=10^{-2}M_{\text{SMBH}}$, $\lambda_{\text{ULA}}=0.3$ and we take $\Upsilon=0$. The axion particle masses correspond to multiples of mean motion and some of the values can be found in Tables \ref{tab:S2_prim_res} and \ref{tab:S2_sec_res}. }
\label{fig:SMBH_semi_major}
\end{figure}

General orbital behaviour is similar to the one described in Section \ref{sec:toy_model}. Numerical calculations show, for the cases that we examined, that (anti-)resonant behaviour of the radial coordinate can be found both when $2\omega
=(2n+1)\tilde{\Omega} $ (odd resonances) and $2\omega
=2n\tilde{\Omega}$ (even resonances), where $\tilde{\Omega}=2\pi/T$ is orbital mean motion, $T$ is orbital period and $n \in \mathbb{N}$\footnote{Similar ratios are known from resonant phenomena (mean-motion resonances) in celestial mechanics and galactic astronomy \cite{book_murraydermott, binney2011galactic}. This observations demands further investigation.}. Resonances become less pronounced, in absolute terms, with increasing $n$. When $\Upsilon=0$ only odd resonances occur and their shape is Gaussian-like. The ``sign'' of these resonances, i.e. whether they lead to increase or decrease of the oribtal radius as well as the timescales involved, depend on the environment. We also find, as in Section \ref{sec:toy_model}, (non-symmetric) window around dominant resonance inside which oscillations are slightly amplified with respect to the other driving frequencies. These cases can be analytically understood with the help of first-order perturbation theory (Appendix \ref{AppEliptical}). In Fig. \ref{fig:SMBH_semi_major} we show the semi-major axis secular evolution for the first three resonant frequencies as well as one non-resonant and one close to the dominant resonance. Qualitative behaviour of the $e$ and $T$ is similar. In Tables \ref{tab:S2_prim_res} and \ref{tab:S2_sec_res} we list amplitudes of secular changes of $a$ for two dominant resonant frequencies, as well as their sign. Notice that the current observational precision $0.16\%$ is enough to probe almost all the scenarios that we examined during the resonant amplification.

\begin{figure}
\centering
\includegraphics[width=1\columnwidth]{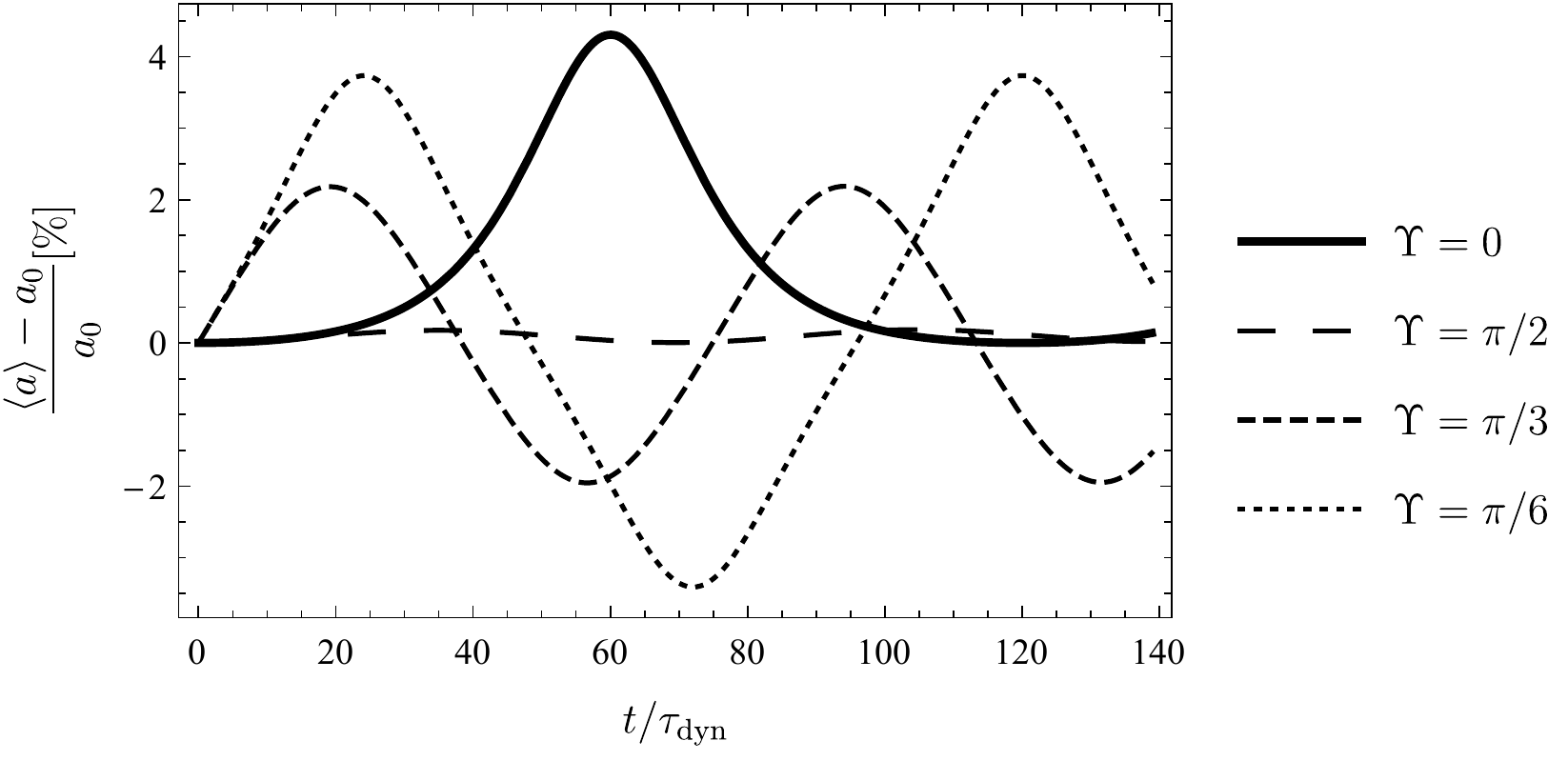}
\caption{Relative change of secular semi-major axis $a$, with respect to the reference one, for dominant (odd) resonant frequency and different values of phase differences $\Upsilon$. Background is the same as in Figure \ref{fig:SMBH_semi_major}.}
\label{fig:SMBH_semi_major_phase}
\end{figure}

\begin{figure}
\centering
\includegraphics[width=1\columnwidth]{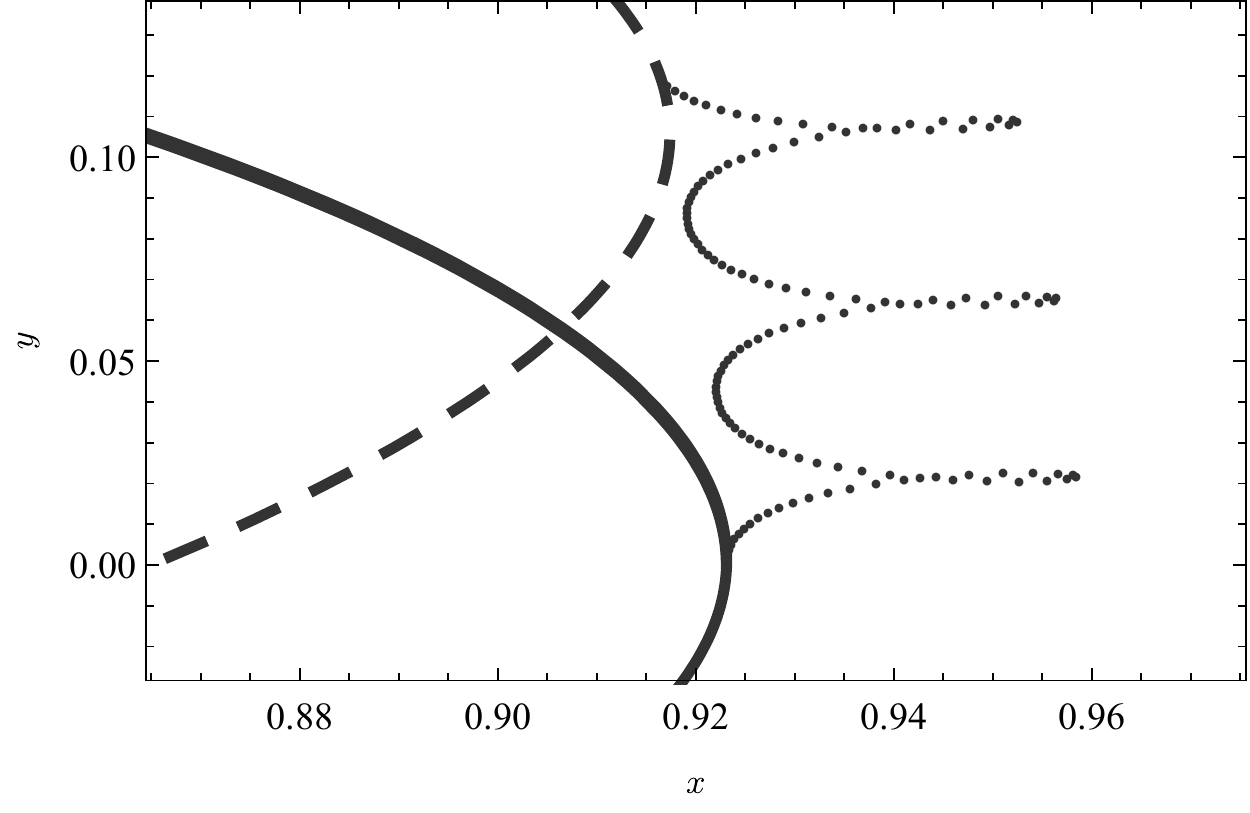}
\caption{Apoastron shift of S0-2 star $(e=0.892)$ during $t=300\tau_{\text{dyn}}$ in orbital plane for $M_\text{ext}=8 \times 10^{-3}M_{\text{SMBH}}$, $\lambda_{\text{ULA}}=0.3$, $\Upsilon=0$ and resonant motion $(2\omega=\tilde{\Omega})$. The orbits are presented at the beginning (full line) and at the end of the interval (dashed). Black dots correspond to the apoastron position during this interval. Notice that SMBH dominates background in determining periastron shift direction (retrograde), but during resonant motion short change of direction of apoastron shift occurs. Orbital coordinates correspond to $x=r\cos{\varphi}$ and $y=r\sin{\varphi}$ and are given in the units of $10 \text{mpc}$.}
\label{fig:SMBH_apoastron}
\end{figure}

For $\Upsilon=\pi/2$ only even resonances occur with the same phenomenology as their odd counterparts. For all other phase differences, of the form $\Upsilon=\pi/m$, $m \in \mathbb{N} \setminus \{1,2\}$, the resonance evolution is akin to a sinusoidal shape (see Fig.~\ref{fig:SMBH_semi_major_phase}). For a given resonance, the amplitude of orbital parameters are mildly dependent on $\Upsilon$: at most (for $a$) they were lowered by $50 \%$ (for $m=3$), compared to the $m=1$ or $m=2$ case, but notice that now $a$ periodically becomes both larger and smaller than its undisturbed value (in absence of time-dependent force). This qualitative dependence on the relative phase is consequence of absence of time-translation symmetry as discussed in Section \ref{sec:symmetries}.
\begin{table}[]
\centering
\caption{Secular amplitudes of S0-2 star semi-major axis $a$, that correspond to primary resonance, and whose sign is denoted by $\sigma$: $+$ for amplification and $-$ for depletion. In each case the homogeneous background mass $M_{\text{ext}}$ (in units of $[0.01 M_{\text{SMBH}}]$) and ULA abundance $\lambda_{\text{ULA}}$ were varied.  We fix phase difference to $\Upsilon=0$. 
Note that the resonance is located at frequencies $\omega=m/\hbar$.}
\label{tab:S2_prim_res}
\begin{tabular}{|c  c | c | c c |}
\hline \hline
$M_{\text{ext}} $  & $m\,[m_{22}]$ & $\lambda_{\text{ULA}}$ & $\sigma$ & $\langle a \rangle \,[\text{mpc}]$   \\
\hline
 
$1$  & $0.0412$ & $0.3$ & $+$  & $5.089$    \\

$$   &  $$ & $0.05$ & $+$  & $4.963$    \\

$$  &  $$ & $0.005$  & $-$  & $4.854$   \\

\hline

$0.1$  & $0.0413$ & $0.3$  & $+$ & $4.931$ \\

$$ & $$  & $0.05$ & $+$ &  $4.895$  \\

\hline

$0$ & $/$ & $0$ & $/$ & $4.878$ \\
\hline
\end{tabular}
\end{table}
\begin{table}[]
\centering
\caption{Same as Table \ref{tab:S2_prim_res} for secondary resonances.}
\label{tab:S2_sec_res}
\begin{tabular}{|c  c | c | c c |}
\hline \hline
$M_{\text{ext}}$  & $m\,[m_{22}]$ & $\lambda_{\text{ULA}}$ & $\sigma$ & $\langle a \rangle \,[\text{mpc}]$   \\
\hline
 
$1$ & $0.124$  & $0.3$ &  $+$  & $4.932$    \\

$$ & $$  & $0.05$  &  $-$  & $4.859$   \\

$0.1$ & $0.124$  & $0.3$  & $+$ & $4.887$  \\
\hline
\end{tabular}
\end{table}
The extended background leads to a retrograde periastron shift of stellar orbits, as reviewed in Section \ref{sec:toy_model}. On the other hand, relativistic effects of the strong SMBH gravity lead to a prograde periastron shift, potentially masking the previous effect. Periastron precession was found by identifying successive radial maximums (in order to evade numerical difficulties explained in Ref. \cite{RubilarEckart:2001}). Sign of $\dot{w}$ direction depends on the background mass $M_{\text{ext}}$ i.e. whether it is dominated by the SMBH PN or background Newtonian contribution. The shape of periastron precession with respect to time is similar to the other orbital parameters. For resonant motion, base value of $\dot{w}$ tends to be lower (in relative terms), as compared to non-resonant one, and $\dot{w}$ increases when amplification occurs. Depending on the values of  $M_{\text{ext}}$ and $\lambda_{\text{ULA}}$ this can lead to change of sign and, as a consequence, direction of periastron precession. This manifests in apoastron developing some kind of helix trajectory as in Fig.~\ref{fig:SMBH_apoastron}. When the oscillating frequency is similar to the resonant one, range of $\dot{w}$ values is similar to the one that correspond to resonant frequency.
For all cases that we considered, value of $\dot{w}$ was inside present constraints.

\begin{figure}
\centering
\includegraphics[width=1\columnwidth]{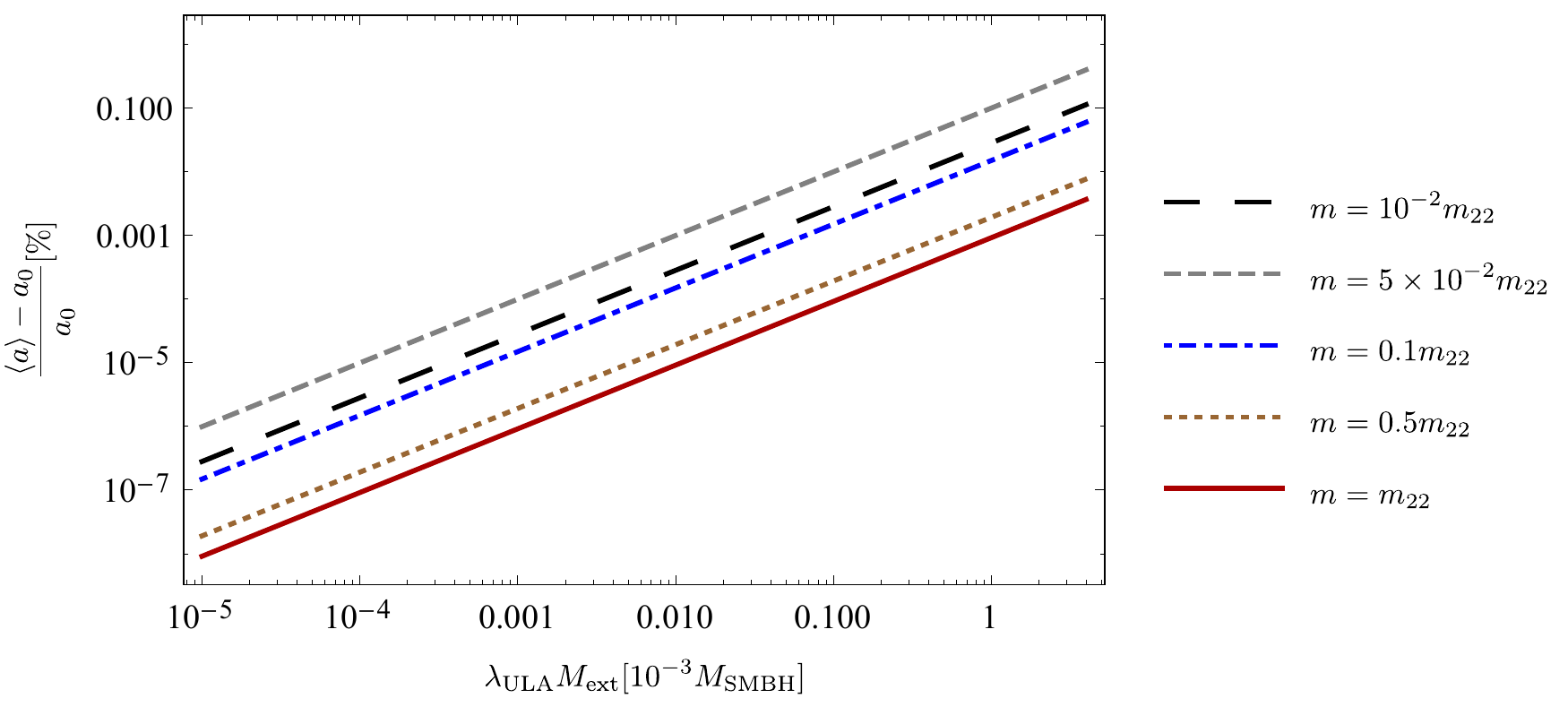}
\caption{Relative change of secular semi-major axis $a$ amplitude, with respect to the reference one for different particle masses and ULA abundances. Results are obtained using first-order pertrubation theory (see Appendix \ref{AppEliptical}) for primary near-resonant window. We have neglected dependence of the phase difference as is it is, in this case, part of the argument of a harmonic function and influences only the time at which  the semi-major axis develops a maximum. Note that the maximum values are obtained for $m=5 \times 10^{-2} m_{22}$ contour, as the primary resonance corresponds to $m=4.1 \times 10^{-2} m_{22}$ (see Table \ref{tab:S2_prim_res}).}
\label{fig:contur_plot_S2_A}
\end{figure}

Inference of ULA mass and abundance from semi-major axis is highly degenerate, as the phase difference, type of resonance and timescales over which the resonance develops significantly contribute to the problem\footnote{Whether inference of other orbital elements could significantly lower the degeneracy of the problem should be subject of further studies.}. A rough picture can be obtained by focusing on a near-resonant window and the first-order perturbation theory, as described in Fig.  \ref{fig:contur_plot_S2_A}. Long-term and precise observations of this and other S stars (and comparison with other constraints) will allow for constraining ULA densities for axion masses that correspond to the resonant frequencies.  Stars with longer orbital periods (or equivalently smaller axion masses), cannot be probed in this way as dynamical timescales become large. However, this type of resonant phenomena is known, in celestial mechanics and galactic astronomy \cite{binney2011galactic, book_murraydermott}, to leave fingerprints in the orbital parameter space, something that deserves further scrutiny. One of the better known of these structures, and potentially similar to this problem, are Kirkwood gaps in the distribution of semi-major axes of the main-belt asteroid orbits \cite{book_murraydermott}. Identification of these structures could be possible with the observation of a large number of stars in the central sub-parsec and parsec scales~\cite{ValluriDebattista:2012}. 
%
\subsection{Spectroscopic fingerprints of halo spacetime} \label{sec:spectro}

In Appendix \ref{AppGredshift} we explore the gravitational redshift induced by oscillaton spacetime oscillation. From the observational perspective, this effect may manifest itself as a systematic error in other spectroscopic measurements, such as radial velocity (RV) curve determination~\cite{2caveat}. For the S0-2 star, luckily data-reduction within a similar context was addressed already: the presence of a binary companion of S0-2 would induce periodic signature in RV measurements. In Ref.~\cite{ChuDoHees:2018} RV residual was formed by subtracting orbital variation of radial velocity. On this residual Lomb-Scargle periodogram was performed with periods sampled between $1$ and $150$ days, which corresponds to $m$ between $4 \times 10^{-2} m_{22}$ and $0.63m_{22}$. Results showed that no significant periodic signal was found. Note that average uncertainties for RV points are $\sim (30-50) \text{km}/\text{s}$. Estimates suggest that with Extremely Large Telescopes (ELT) such uncertainties will be lowered by an order of magnitude and a bit lower \cite{HeesProc:2017}.

Taking the MW estimates from Section~\ref{sec:darkhalo_descr}, we find that the gravitational redshift (modulation) amplitude is $z \sim 10^{-6}$ (see Appendix~\ref{AppGredshift}). This would lead to a periodic change of $\sim 0.1 \text{km}/\text{s}$, corresponding to a mass $m \sim 0.8m_{22}$. This value is smaller than the expected uncertainties with ELT. The presence of baryons, which contribute to the static metric component, would make this amplitude even lower.

\section{Discussion}

The motion of objects in time-period spacetimes has a number of distinctive features. Suprisingly,
and despite the potential application to dark matter physics at the galactic center, most of these features had not been discussed previously.
We have shown that the periodic forcing on orbiting stars and planets, exherted by oscillating dark matter leads to unique
features in the motion of such objects. When applied to stars close to the galactic center, such features may well be observed in the near future.
Together with resonances excited by {\it non-axisymmetric} ultralight dark matter close to BHs~\cite{Ferreira:2017pth,Brito:2014wla,Fujita:2016yav,Brito:2017zvb,Cardoso:2018tly,Baumann:2018vus},
or by homogeneous oscillating DM in binary systems~\cite{Khmelnitsky:2013lxt}, our results suggest very strongly that the next few years will see either strong constraints on such DM models, 
or very strong indications that it describes our universe.

\begin{acknowledgments}
We thank Philippe Grandclement for useful correspondence and for sharing with us his spectral code to build oscillatons, which we used to check our own
numerics. We would like to thank the anonymous reviewer for useful comments. M.B. acknowledges discussions on galactic phenomenology with Nemanja Martinović and the generous support of the GWverse COST Action CA16104 on ``Black holes, gravitational waves and fundamental physics''.
M. F. acknowledges financial support provided by Funda\c{c}\~{a}o para a Ci\^{e}ncia e a Tecnologia Grant number PD/BD/113481/2015 awarded in the framework of the Doctoral Programme IDPASC - Portugal.
The authors acknowledge financial support provided under the European Union's H2020 ERC Consolidator Grant ``Matter and strong-field gravity: New frontiers in Einstein's theory'' grant agreement no. MaGRaTh--646597. Research at Perimeter Institute is supported by the Government of Canada through Industry Canada and by the Province of Ontario through the Ministry of Economic Development $\&$
Innovation.
This project has received funding from the European Union's Horizon 2020 research and innovation programme under the Marie Sklodowska-Curie grant agreement No 690904.
\end{acknowledgments}

\appendix
\section{Weak field limit of the Einstein-Klein-Gordon equations for the real scalar field} \label{AppEKGNewt}
The small compactness regime of oscillatons and boson stars corresponds to the Newtonian-like limit: velocities are small, and the gravitational
potential is everywhere weak. We then write the truncated metric coefficients corresponding to the Einstein-Klein-Gordon background, \eqref{EKGmetricB} and \eqref{EKGmetricA}, as slightly perturbed away from the Minkowski metric:
\beq
A(t,r) &=& 1 + 2V(r)+2V_{1}(r) \cos(2 \omega t)\,,\label{eq:NexpA2}\\
B(t,r) &=& 1 + 2W(r)+ 2W_{1}(r) \cos(2 \omega t)\,.\label{eq:NexpB2}
\eeq
The ansatz for the field \eqref{EKGmetricPhi} is
\be 
\Phi(t,r) = \phi(r) \cos(\omega t) \,.\label{eq:NexpPhi2}
\ee
As noted in Section~\ref{sec:NOscillatons}, we are expanding the frequency of the scalar field around its mass so that, up to the second order in wave number $k$, 
we can write $\omega=\mu+k^2/(2\mu)+{\cal O}(k^4)$. Our expansion parameter is the group velocity $v=k/\mu$.

The order of magnitude of the various derivatives in the weak field regime can be estimated using $\phi_{1}(r) =e^{ikr}$. Then: 
\beq 
&&\partial_t \Phi(t,r) \sim - \mu \phi(r) \sin(\omega  t)+{\cal O}(v^2) \label{eq:NtderivativePhi} \,,\\
&& \Phi'(t,r) \sim {\cal O}(v)\,.  \label{eq:NrderivativePhi}
\eeq
In the Newtonian limit of the Einstein's equations we expect that $V \sim 1/r \sim {\cal O}(v^2) $ and $W \sim {\cal O}(v^4) $ and mutatis mutandis for $V_{1}$ and $W_{1}$. Differentiating \eqref{eq:NexpA} with respect to the time and radial coordinates we obtain:
\beq
&& A'(t,r) \sim {\cal O}(v^4) \,, \label{eq:Nrderivativea} \\
&&\partial_t A(t,r) =- 2\mu V_1(r)\sin(2 \omega t) + {\cal O}(v^4) \,. \label{eq:NtderivativeA}
\eeq
%

Applying this ansatz to \eqref{eq:4of4} we obtain \eqref{eq:NSchrodinger}
\be
e\phi=-\frac{1}{2\mu^2 r}(r\phi)''+V\phi\,,
\ee
with $e=k^2/2\mu$. 

In order to get Poisson equation, we will follow the approach of obtaining weak-field limit of a relativistic star \cite{bookWeinberg:1972}. We will introduce notation $\nu=2V(r)+2V_{1}(r)\cos(2 \omega t)$ and $\sigma=2W(r)+2W_{1}(r)\cos(2 \omega t)$. Einstein tensors $G_{tt}$ and $G_{rr}$ are:
\beq
G_{tt}&=&\frac{A}{r^2} \Big( \frac{1}{B} \Big( \frac{B'}{B}r-1   \Big)+1\Big)\,,\\
G_{rr}&=&\frac{B}{r^2} \Big( \frac{1}{B} \Big( \frac{A'}{A}r+1   \Big)-1\Big)\,.
\eeq
Using \eqref{eq:NexpA} and \eqref{eq:NexpB} and expanding up to $v^4$ we get:
\beq
G_{tt}r^2&=&\sigma+ r \sigma'\,, \\
G_{rr}r^2&=&r\nu' -\sigma\,.
\eeq
Equating last two expressions with the corresponding component of the stress-energy tensor \eqref{eq:stressenergy} we obtain:
\beq
(\sigma r)'&=& 8\pi r^2T_{tt} \,,\label{eq:NEinsteintt}\\ 
\nu' &=& 8\pi rT_{rr}+\frac{\sigma r}{r^2} \,.  \label{eq:NEinsteinrr}
\eeq
Differentiating equation \eqref{eq:NEinsteinrr} and combining it with \eqref{eq:NEinsteintt}
\be
\nu''+\frac{2\nu'}{r}=8\pi ( r T'_{rr}+T_{rr}+T_{tt})\,.\label{eq:NalmostPoisson}
\ee
If we reinstate $c$ and remember that $G_{\mu \nu}=(8\pi/c^4)T_{\mu \nu}$, it stands out that the term in brackets on the right hand side of the last equation should be treated up to $c^2$ order i.e.
\be
rT'_{rr}+T_{rr}+T_{tt}=\frac{1}{2}\phi^2\mu^2-\cos(2\omega t)\left(\frac{1}{2}\phi^2\mu^2+r\mu^2\phi \phi' \right)\,.
\ee
If we redefine the scalar, $\phi = \psi/\sqrt{8 \pi}$ (as in Section \ref{sec:Oscillatons}), and equate terms in front of the $\cos(0)$ and $\cos(2\omega t)$ on both sides of the equations we find:
\beq
&&V''+\frac{2V'}{r}=\frac{1}{2} \psi^2\mu^2 \,,\label{eq:NPoissonApp}\\ 
&&V''_{1}+\frac{2V'_{1}}{r}=-\frac{1}{2} \psi^2\mu^2-r\mu^2\psi \psi'\,.\label{eq:NTimePoisson}
\eeq
Equation \eqref{eq:NPoissonApp} is nothing but Poisson equation. We re-write \eqref{eq:NTimePoisson} as,
\be
V_{1}'= -\frac{1}{2} r\psi^2\mu^2+\frac{1}{2r}\int^{r}_{0} r^2\psi^2\mu^2 dr\,.\label{eq:NTimePoisson2}
\ee
We see from \eqref{eq:NEinsteinrr} that $\sigma \sim \partial_{r}{\nu} \sim {\cal O}(v^4)$ as claimed. Finally, using \eqref{eq:NEinsteintt} we see that the second term on the r.h.s of \eqref{eq:NTimePoisson2} is of order $o(v^6)$. Thus, the second term (mass) on the r.h.s. of \eqref{eq:NTimePoisson2} is smaller than the first and can be neglected.
Therefore, setting $\mu=1$, we finally obtain equations \eqref{eq:NPoisson} and \eqref{eq:NV2}:
\beq
(rV)''&=&\frac{1}{2} r\psi^2\,,\\
V'_{1}&=&-\frac{1}{2} r\psi^2\,.
\eeq
%

\section{Scalar field DM models} \label{AppGPP}

As mentioned in the Introduction, various types of scalar fields have been considered as DM models. Most of them develop cores that can be described by the Gross-Pitaevskii-Poisson (GPP) system \cite{Berezhiani:2015bqa, Hui:2016ltb, RindlerDallerShapiro:2012}. In the other extreme, when the self-gravitating scalar configurations are candidates for compact objects, GPP is a useful arena for understanding qualitative aspects of models. The GPP equation describes systems of large number of bosons at zero temperature, with the mean-field approximated self-interaction and coupled to the external potential \cite{RogelSalazar:2013}
\be
e\psi=\left(-\frac{\hbar^2}{2\mu}\nabla^2+V_{\text{ext}}+U(\psi)\right)\psi\,.
\ee
Usually, self-interactions $U(\psi)$ are described up to two-body processes $g_{2}|\psi|^2$. The external potential is gravitational, in the DM physics context, and various optical/magnetic traps in the cold atoms research. The gravitational potential $V_{\text{ext}}=\mu V$ is described by the Poisson equation. When the self-interaction is negligible the system is called Schrodinger-Poisson (SP). It should be noted that we have chosen a normalization of the wavefunction in the form $\int dV |\psi(r)|^2=N$, where $N$ is the number of the particles in the system.

FDM models presume very small attractive self-interaction that can be neglected on a galactic scale \cite{Hui:2016ltb,Desjacques:2017fmf}. On the other hand, scalar field DM models with strong repulsive\footnote{Labeled as a ``wrong sign'' potential in the context of axion stars.} self-interaction has also been considered~\cite{RindlerDallerShapiro:2012, Chavanis:2011}. In the superfluid DM, the condensate (when the phonon excitations are neglected) is characterized by the self-interaction through primarily three-body processes, captured in the GPP picture by $g_{3}|\psi|^4$ \cite{Berezhiani:2015bqa}. Other types of self-interactions have also been studied \cite{Macedo:2013jja,UrenaLopez:2001tw}. Self-interactions are important in the context of ultra-compact objects~\cite{Cardoso:2017cqb,Cardoso:2016oxy,Helfer:2017a}.

Gross-Pitaevskii equation can, in a hydrodynamical picture, be rewritten as Navier-Stokes equation \cite{RogelSalazar:2013, Chavanis:2011} with the present additional term $(1/m)\nabla Q$, called quantum pressure\footnote{Although the more appropriate label would be quantum potential as it is not in the form $(1/\rho) \nabla p^{(Q)}$. Term can be expressed as $(1/\rho) \del_i {p^{(Q)}_{ij}}$\cite{Mocz:2017}.} $Q$ and defined as
\begin{equation}
\frac{1}{m} Q=-\frac{1}{2\mu^2} \frac{\nabla^2 \sqrt{\rho}}{\sqrt{\rho}}\,.
\end{equation}
This term originates from the kinetic term in the Hamiltonian. Pressure arising from (repulsive) self-interaction of both dominant two- and three-particle processes, has a polytropic equation of state 
\be
P=K\rho^{1+1/n}\,,
\ee
where $K=g_{2}/(2\mu^2)$ ($K=2g_{3}/(3\mu^3)$) for dominant two (three)-particle  interaction. When the flow is stationary and the quantum pressure is negligible with respect to the the pressure originating from the (repulsive) self-interaction and gravity, system is in the Thomas-Fermi regime and Navier-Stokes and Poisson equation reduce to Lane-Emden equation with polytropic indices $n=0.5$ for superfluid DM and $n=1$ for repulsive/fluid DM. 
 
In the rest of this Appendix we will describe details of our analytical construction of the profile of the Newtonian oscillatons. Detailed analytical and numerical analysis of the SP/GPP system can be found elsewhere, e.g. \cite{Hui:2016ltb,Chavanis:2011,Guzman:2004wj,KlingRajaraman:2017, KlingRajaraman:2017b} and references therein.
\subsection{Analytical profile of non-self-interacting Newtonian oscillatons}  \label{AppGPP_non_self}
As described in Section \ref{sec:NOscillatons}, the SP equation admits a scaling symmetry, given by \eqref{eq:NSPscale}, so that solutions corresponding to different masses can be obtained from a unique solution by rescaling. This symmetry is preserved even with an addition of self-interacting term \cite{KlingRajaraman:2017b}.

When the self-interaction is negligible, one can find an approximate, analytic solutions of the SP system~\cite{KlingRajaraman:2017}. The scale-invariant field $s=\psi /\lambda^2 $, where $\lambda=\sqrt{\mathcal{C} Z/\beta}$ (scale factor), is found by expanding around zero value of the radial coordinate and at infinity and matching these solutions. Once the field is known, density can be found as $\rho=\mu^2\psi^2/(8\pi)= \Lambda (
\mu s)^2$ and $\Lambda=\lambda^4/(8\pi)$. Expansion of the scale-invariant field around the center is given by
\beq
s_{<}=\sum^{\infty}_{n=0}s_{n}z^n, \label{eq_appB_small_r_exp} 
\eeq
where $z$ is the scale-invariant radial coordinate $z=\lambda \mu r$. This expansion is not convergent after $z>4$. At large radius adequate expansion is of the form
\beq 
s_{>}=\sum^{\infty,\infty}_{n,m=0,0} s^n_{m}\Big(\frac{e^{-z}}{z^\sigma} \Big)^n z^{-m}. \label{eq_appB_large_r_exp}
\eeq
The series in $m$ is only asymptotic to the $s$, for large $z$, and in Ref. \cite{KlingRajaraman:2017} optimal asymptotic approximation \cite{benderbook} is performed by truncating the series with the adequate $M$. From this expansion we see that the long-range behaviour of the density is
\be
\rho(r) \sim \Lambda \mu^2 \alpha^2 (\lambda \mu r)^{2\sigma} e^{-2 \lambda \mu  r} \,,
\ee
where $\alpha=s^1_0$, $\sigma=1+\beta$.

Object linearly related to the scale-invariant Newtonian gravitational potential is defined as
\be
2\Big (\frac{e}{\mu} - V \Big )=\lambda^2 v \,.
\ee
Expansions for $v$ have the same form as for $s$:
\beq 
v_{<}=\sum^{\infty}_{n=0}v_{n}z^n \,,\, v_{>}=\sum^{\infty,\infty}_{n,m=0,0} v^n_{m}\Big(\frac{e^{-z}}{z^\sigma} \Big)^n z^{-m}. \label{eq_appB_v_exp}
\eeq
Series coefficients can be found by inserting expansions for $s$ and $v$ into SP system \cite{KlingRajaraman:2017}. Then, the expansions are matched at the matching point $2.5<z_{\star}<3.5$ and free parameters $s_{0}$, $\alpha$, $\beta$ and $v_{0}$ are found by fitting onto numerically obtained solutions. For the parameter values, we used one given in Eq. 31 in Ref. \cite{KlingRajaraman:2017} and reconstructed terms up to $N=50$ for $s_{(N)} \approx  s_{<}$ and $N=3,M=6$ for $s^{(N)}_{(M)} \approx s_{>}$, where $N$ and $M$ refer to orders of series truncation, with $z_{\star}=3$. Value of the scale-invariant radius $Z$ is found by inverting $m(Z)=0.98 M$.

\subsection{Analytical profile of self-interacting Newtonian oscillatons} \label{AppGPP_self_int}

We focus here only on two-particle interactions,  $U(\psi)=g_{2}|\psi|^2$ in the GPP system. It is well known that the Lane-Emden equation with $n=1$ allows for analytical solution of the form $\rho=\rho_{c}\sin{(\xi)}/(\xi)$, where $\rho_{c}$ is density at the center of the polytrope, $\xi=r/\alpha$ is a standard notation for the dimensionless radius of polytropes and $\alpha^2=(n+1)K\rho_{c}^{1/n-1}/(4\pi)$ is the scale factor. Polytropes with $n<5$ do admitt well-defined surface and their radius-mass relation is given by 
\begin{equation}
R \propto M^{\frac{1-n}{3-n}},
\end{equation}
where the constant of proportionality depends on $K$ and $n$ \cite{poissonwillbook}.

Recently, the methodology of the approximate analytical solution construction for the SP system that we have used in this paper \cite{KlingRajaraman:2017}, has been expanded to include self-interactions \cite{KlingRajaraman:2017b}. For weak couplings, the expansion is perturbatively constructed around non-self-interacting solution. Intermediate and strong coupling are obtained by perturbations around the Thomas-Fermi solution, described in the previous paragraph.

We parametrise the strength of the self-interaction, as in Ref. \cite{KlingRajaraman:2017b},
\begin{equation}
\gamma=\frac{\Delta \Lambda(\gamma)}{2\mu^2}.
\end{equation}
Fixing the value of $\gamma>\gamma_{\text{min}}$ and $\mathcal{C}$ uniquely specifies the value of $\Delta=32\pi \mu g_2$, as can be seen on Fig. 4 in Ref. \cite{KlingRajaraman:2017b}. There is a minimum value of coupling for attractive self-interactions $\gamma_{\text{min}}=-0.722$, after configurations become unstable under small perturbations and hence unphysical \cite{KlingRajaraman:2017b}.

For weak couplings, the field and potential expansion is the same as in \eqref{eq_appB_small_r_exp}, \eqref{eq_appB_large_r_exp} and \eqref{eq_appB_v_exp}, but now free parameters depend on the strength of the self-interaction (Eq. 29 in Ref. \cite{KlingRajaraman:2017b}). As we are interested only in general aspects of this models, we used $N=20$ for $s_{(N)}$ and $s^{(N)}_{(M)}$ is identified with the Whittaker function, that well describes leading order long-range behaviour \cite{KlingRajaraman:2017,KlingRajaraman:2017b}. The matching point was estimated as a point where difference between $s_{(N)}$ and $s^{(N)}_{(M)}$ is the smallest.
For the strong coupling we used Thomas-Fermi solution, matched with the Whittaker function (for the long-range behaviour). The matching point can be found as the point where the self-interacting and the quantum pressure have the same value \cite{KlingRajaraman:2017b}.

\section{Motion in self-interacting Newtonian oscillatons} \label{AppGPP_self_int_motion}

As we have seen in Section \ref{sec:NOscillatons}, the description of the time-dependent potential \eqref{eq:NV2} is decoupled from the SP/GPP system. In this way, the profile description of Appendix \ref{AppGPP} is valid for both complex and real scalar self-gravitating configurations. A time-dependent potential exists only in the second case \cite{UrenaLopez:2002gx}. 

We will now find the ratio of the orbital frequency and the frequency of oscillatons in the weak coupling and in the Thomas-Fermi regime. The qualitative picture is simple: attractive self-interactions make the object more compact. This causes $\tilde{\Omega}(z)/\omega$ to increase, with respect to the non-self-interacting value. On the other hand, the time-dependent potential will be overtaken by the Newtonian gravitational potential at smaller distances. Repulsive self-interaction will make an opposite effect. From scaling relations (and $\omega \approx \mu$), we find
\begin{equation} \label{eq:osc_self_int_omega_ratio}
\frac{\tilde{\Omega}(z)}{\omega} = \mathcal{C}\frac{Z(\gamma)}{\sqrt{2}\beta(\gamma)}\sqrt{\frac{1}{z^3}\int^{z}_{0}y^2(s(y;\gamma))^2 dy}.
\end{equation}
The ratio is proportional to $\mathcal{C}$, with a coefficient of order one, see Fig. \ref{fig:res_self_int_osc} (where we used the profile description given in Appendix \ref{AppGPP} to compute the integral). As $\mathcal{C} \leqsim 0.01$ in the weak-field regime, even with self-interactions included, motion in the oscillatons will be in the rapidly oscilating background regime. 
\begin{figure}
\centering
\includegraphics[width=0.5\textwidth]{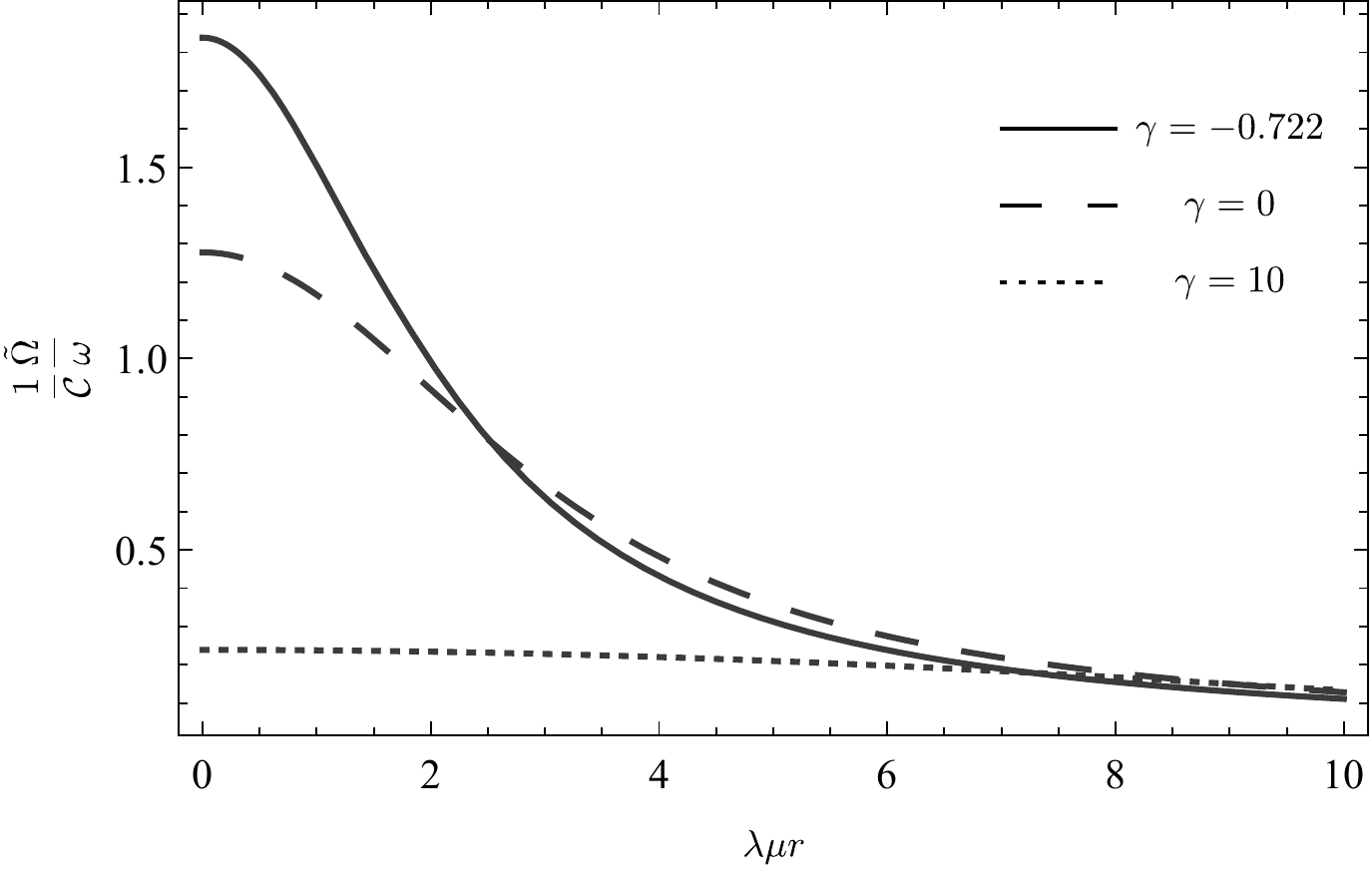}
\caption{Ratio of Keplerian and oscillaton frequency for different signs and values of the scalar self-interacting coupling}
\label{fig:res_self_int_osc}
\end{figure}
%
\section{Elliptical orbits in ultralight DM background: perturbation theory} \label{AppEliptical}

In order to analytically understand the influence of time-periodic backgrounds on elliptical orbits in the context of ULA DM, consider first-order perturbation theory in celestial mechanics \cite{poissonwillbook}. We here assume that some other matter components, e.g. SMBH as in Section \ref{sec:darkhalo_res}, are dominant and consider DM background as a homogeneous one. This approach is equivalent to the one in Ref. \cite{Blas:2016ddr}, but stated in a different language. For example, taking secular equation for the change of semi-major axis $a$ and working at the first order we obtain
\begin{equation}
\langle \dot{a} \rangle_{\text{ULA}}(t) =-4\rho_{\text{ULA}}T a \frac{J_n (ne)}{n}f(t),
\end{equation}
where we used notation from Section \ref{sec:darkhalo_res}, $J_n$ is Bessel function, $f(t)=\sin(\delta \omega t+ 2\omega t_{0}+2\Upsilon)$, $\delta \omega=2\omega-n\tilde{\Omega}$ and $t_0$ is the time of the first periastron passage\footnote{In order to obtain this equation, one should change, when averaging, from integrating with respect to time to integrating with respect to the eccentric anomaly and use Kepler's equation. Also, we assumed that $\delta \omega \ll 1$ as the term $\delta \omega t$ is approximately constant on the orbital timescales.}. Prediction of this equation is in good agreement with our numerical results from Section \ref{sec:darkhalo_res} for the driving frequencies close to the resonante one. Note that as we work at the first order, we can add different contribution to orbital element secular change. In the context of Section \ref{sec:darkhalo_res}:  $\langle a \rangle=\langle a \rangle_{\text{ULA}}$, as $\langle a \rangle_{\text{PN}}=0$.

This result is \textit{not} applicable when the true resonance occurs  i.e. $\delta \omega$ is sufficiently small as it would imply linearly diverging secular evolution of $\langle a \rangle$. Secular evolution is tamed at some point in time, as observed in Section \ref{sec:darkhalo_res}, and this phenomenon is the generic one as discussed in Sections \ref{sec:circularmotionexample}-\ref{sec:toy_model}. 

\section{Motion in stellar and planetary systems inside ultralight DM halo} \label{AppBinary}

Consider now how time-periodic forces influence the motion of objects within a stellar or planetary system, which itself moves around a halo. For simplicity we focus on binary systems, like those studied recently~\cite{Khmelnitsky:2013lxt, Blas:2016ddr,DeMartinoBroadhurst:2017}. Objects in such systems experience acceleration (neglecting all other contributions except their mutual gravitational interaction and time-dependent force)
\begin{equation}
m_{i}\vec{r}_{i}=\pm\frac{m_{1}m_{2}}{r^3}\vec{r}+\vec{F}^{\text{TD}}_{i}\,,
\end{equation}
where $\vec{F}^{\text{TD}}_{i}=4\pi \rho(\vec{r}_{i}) m_{i} \vec{r}_i \cos{(2\omega t+2\Upsilon(\vec{r}_{i}))}$ is the time-dependent force in the context of ULA DM and $\vec{r}=\vec{r}_2-\vec{r}_1$. We can transform perspective to that of dynamics of the relative particle $m_{\text{red}}=m_1 m_2 /M$, where $M=m_1+m_2$, and the center-of-mass (CM) $\vec{R}=(m_1 \vec{r}_1 + m_2 \vec{r}_2)/M$. Note that $\rho(\vec{r}_i) \approx \rho(\vec{R})$ and $\Upsilon(\vec{r}_i) \approx \Upsilon(\vec{R})$. Thus, time-dependent force decomposes to center-of-mass and relative component. Relative component is just a small perturbation with respect to the other forces that dictate dynamics in stellar and planetary systems, but it can lead to potentially observable consequences if resonance occurs. For example, Ref.~\cite{Blas:2016ddr} studied how it influences the secular change of binary pulsar periods and whether such effect is measurable. As noted in Ref.~\cite{Pani:2015qhr} there are two more contributions to the period change, that of the CM motion along the line-of-sight and the one induced by variation of orbital inclinations. The last effect depends on the binary orbit orientation. Furthermore, Solar System barycenter, with respect to whom pulsar timing are measured, also oscillates. 

Let us estimate ULA DM consequence on relative motion in the Solar System. Using first order perturbation theory (see Appendix \ref{AppEliptical} and Ref. \cite{Blas:2016ddr})
\beq
\langle \dot{a} \rangle_{\text{ULA}}(t) &\approx& -3 \times 10^{-5} \frac{\text{m}}{\text{yr}} \Big ( \frac{\rho_{\text{ULA}}}{1.13 \times 10^{-2} \frac{M_\odot}{\text{pc}^3}}  \Big )\nonumber\\
&& \Big ( \frac{T}{1\text{yr}} \Big ) \Big ( \frac{a}{1\text{AU}} \Big ) \frac{J_n (ne)}{n}f(t)\,.
\eeq
We used values for the local DM density from Ref.~\cite{SalucciNesti:2013}. Present accuracy in Solar System observations is $\sim 10^{-1} \text{m}/\text{yr}$ for Mars and $\sim 10^{-2} \text{m}/\text{yr}$  for Moon \cite{Minazzoli:2017vhs}. Note also that Solar System planets have low eccentricites and $J_n$ additionally suppresses this ratio: e.g. for Mars $e=0.0934$ and $J_1(e)\approx4\times10^{-2}$. We stress that the truly resonant case should be properly studied numerically.

\section{Gravitational redshift in time-dependent geometries: oscillaton spacetimes} \label{AppGredshift}

Observers measure different proper-time intervals depending on the properties of the underlying metric; thus, two observers may measure different frequencies for the same pulse of radiation 
if they are at different points in spacetime. This frequency difference can be quantified by the quantity
\begin{equation}
\frac{\omega_i - \omega_j}{\omega_j}\,,
\end{equation}
where $\omega_i$ and $\omega_j$ are frequency measurements (of the same radiation signal) by two different observers. This quantity is called the gravitational redshift.

Given the umbilical relation between the spacetime and the gravitational redshift, for a time-dependent metric the redshift also varies in time. 
This possibility has been investigated, in a non-cosmological context, for light propagating through a gravitational wave~\cite{kaufmann1970redshift,faraoni1991frequency,smarr1983gravitational} and in the context of quantum fluctuations of spacetime \cite{thompson2006spectral}. For simplicity, but without loss of generality, focus on radiation emitted by a star at rest at the origin $r=R_e=0$ and received at a distance $r=R_r$, in an oscillaton spacetime.
Considering that the emitter and the receiver are at rest, one can write their 4-velocities as
\begin{equation}
u^{\mu} = \left(\frac{1}{\sqrt{A(t,r)}},0,0,0\right),
\end{equation}
such that the 4-momentum of a radial photon is
\begin{equation}
P^{\mu} = \left(\frac{E_r}{\sqrt{A(t,r)}},\frac{E_r}{\sqrt{B(t,r)}} ,0,0\right)\,,
\end{equation}
where $r=R_e,\,R_r$ is the radial coordinate of emission or of reception of the signal. 
Notice that these expressions satisfy the definition of the 4-momentum of a photon~\cite{bookWeinberg:1972,poissonwillbook},
\begin{equation}
P^{\mu}P_{\mu} = 0 \quad \text{and} \quad -E_r = P^{\mu}u_{\mu}.
\end{equation}

The energy of the photon is given by $E_r = \hbar \omega_r$, where $\omega_r$ is the associated frequency; using this definition together with the explicit form of its 4-momentum, one can write the redshift between the emitter and receiver as
\begin{equation}
\label{eq:red-general}
Z\equiv 1 + z \equiv \frac{\omega_e}{\omega_r} = \sqrt{\frac{A(t_e,R_e)}{A(t_r,R_r)}}\frac{P^t(R_e)}{P^t(R_r)}\,,
\end{equation}
where $P^t$ stands for the time component of the 4-momentum of the photon and $R_e$ and $R_r$ correspond to the radial coordinates of the emitter and the receiver of the photon, respectively. This expression can be calculated by numerically solving the geodesic equation for the 4-momentum of the photon
\begin{equation}
\frac{dP^{\mu}}{d\lambda} + \Gamma^{\mu}_{\alpha \beta} P^{\alpha} P^{\beta} = 0,
\end{equation}
where the Christoffel symbols $\Gamma^{\mu}_{\alpha \beta}$ are calculated using the full time dependent metric of Eq.~\eqref{metric_expansion}.

For static spacetimes, there's an exact time-like Killing vector which guarantees that $P^t(R_r) a_0(R_r) = P^t(R_e) a_0(R_e)$. Using this result in Eq.~\eqref{eq:red-general} it follows that
for static spacetimes,
\begin{equation}
\label{eq:static-redshift}
Z_{\rm static} = \sqrt{\frac{a_0(R_e)}{a_0(R_r)}} \frac{a_0(R_r)}{a_0(R_e)} = \sqrt{\frac{a_0(R_r)}{a_0(R_e)}}\,,
\end{equation}
a well-known result~\cite{schunck1997gravitational,perlick2004gravitational}.

\begin{figure}
\centering
\includegraphics[width=1\columnwidth]{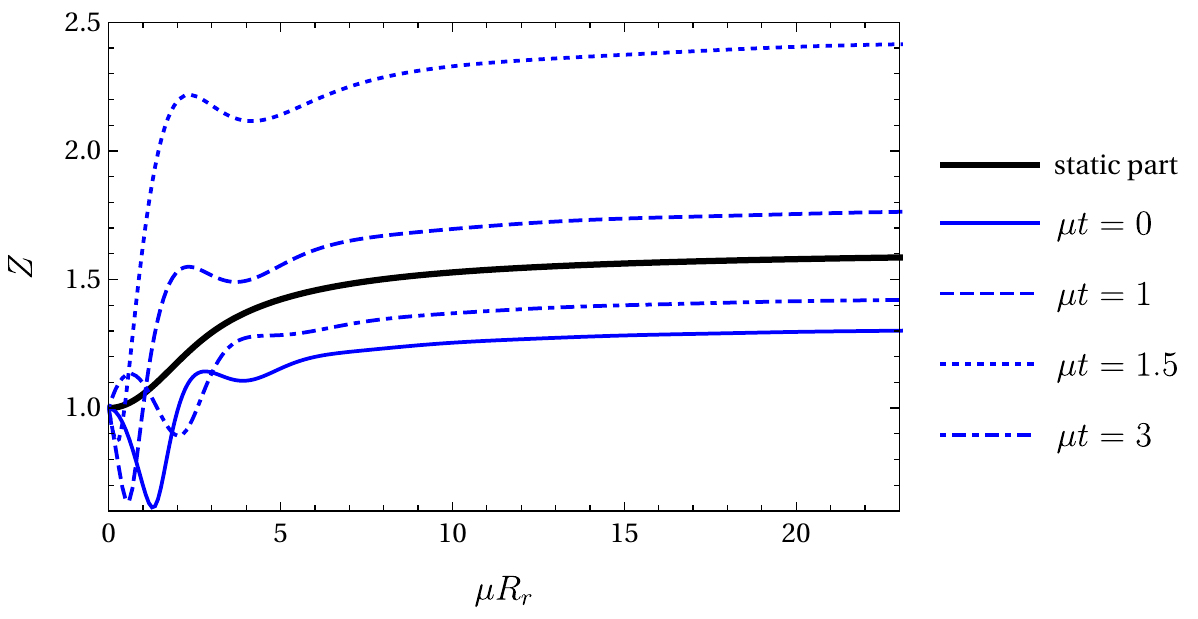}
\caption{Variation of the redshift of photons emitted at $R_e=0$ as a function of the position $R_r$ of the receiver; the underlying spacetime is generated by an oscillaton with $\Psi_1(0)=0.35$ ($\mathcal{C} \sim 0.07$). The solid black line represents the case in which the time-dependent part of the metric is neglected. The blue lines, with several patterns, were calculated using the full metric and correspond to photons emitted at different instants of coordinate time. The value of the coordinate time is not important, what is relevant is the ordering that the time coordinate provides: two consecutive photons will be redshifted differently depending on the amount of coordinate time that separates their emission event. Notice the formation of a plateau as the value of $Z$ stabilizes for larger values of $R_r$; this stabilization occurs as $R_r$ increases past the value of the radius of the oscillaton.}
\label{fig:RvsR}
\end{figure}

The relation in Eq.~\eqref{eq:static-redshift} can be used as a good measure of comparison for the effects of the time-dependent part of the metric. Most notably, the corresponding relation for the full metric, Eq.~\eqref{metric_expansion}, will be a function of the coordinate time, as can be seen in Fig.~\ref{fig:RvsR}. The different curves in this figure show that the time variation of the metric influences the value of the redshift through all the radial extent that separates the source (in the case of Fig.~\ref{fig:RvsR} at $R_e = 0$) and the receiver. Moreover, it is evident that a continuous flow of monochromatic photons will be received at $R_r$ with a time dependent frequency.

\begin{figure}
\centering
\includegraphics[width=1\columnwidth]{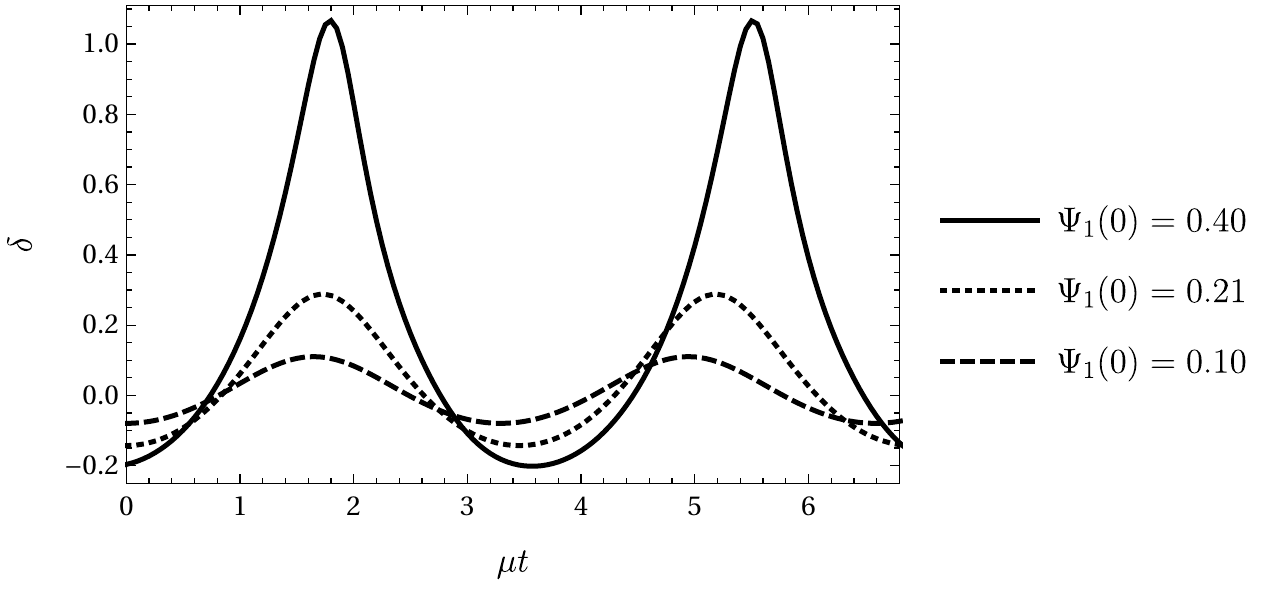}
\caption{Variation in time of redshift measured at large $R_r$, i.e., measured by an observer in the spatially flat part of the metric; the emitter of radiation is placed at $R_e = 0$. Each line corresponds to different oscillaton configurations. The magnitude of the redshift variation grows with the central value of the scalar field $\Psi_1(0)$. The redshift is modulated with a frequency equal to the fundamental frequency of the corresponding oscillaton.}
\label{fig:RvsT}
\end{figure}

A given observer can use the gravitational redshift of a spectral line to compute the gravitational potential where the source was located.
The gravitational redshift in periodic spacetimes will be periodic, and both the period and amplitude of this modulation can inform the observer on the type of time-periodic spacetime.
It is useful to define
\be
\delta=\frac{Z-Z_{\rm static}}{Z_{\rm static}}\,.
\ee
This quantity is shown, at a fixed reception point, in Fig.~\ref{fig:RvsT}. The redshift variation is clearly periodic with a period $\pi/\omega$.
where $\omega$ is the fundamental frequency of the oscillatons, Eqs.~\eqref{eq:expB}, \eqref{eq:expC}, \eqref{eq:expPhi}. 
Thus, if the amplitude of such oscillations is large enough, the oscillaton fundamental freqency can be directly inferred. The amplitude $|\delta_{\text{max}}|$, increases with the magnitude of the central value of the scalar field, as might be expected; numerically, we find that
\begin{equation}
|\delta_{\text{max}}| \sim 1.2 \Psi_1(0) + 37\Psi_1(0)^{4.5}\,,
\end{equation}
a relation that is valid for oscillatons with $\Psi_1(0) < 0.4$. One can also relate $|\delta_{\text{max}}|$ with the relative magnitude of the $g_{tt}$ functions at the origin, i.e, $\varepsilon = a_1(0)/a_0(0)$; we find that
\begin{equation}
 |\delta_{\text{max}}| \sim 0.6 \varepsilon + 0.8 \varepsilon^{3.8},
\end{equation}
valid for $\varepsilon < 0.8$. These two expressions were calculated using stable oscillatons only. 
%
Using these fits it can be estimated a value for the maximum redshift variation in the case of a scalar field halo embedding a galaxy [see Eq.~\eqref{eq:compactness_halo}]. For the particular case of the Milky Way, this value is
\begin{equation}
|\delta_{\text{max}}|_{\text{MW}} \sim 1.34 \times 10^{-6}.
\end{equation}

\bibliography{references}

\begin{thebibliography}{105}
\bibitem [{wea()}]{weakly}%
  \BibitemOpen
  \href@noop {} {}\bibinfo {note} {This fact, that there is no \textit{weakly
  dynamical} approximation of the oscillatons was mistakenly interpreted as an
  absence of the \textit{weak field} limit~\cite{Seidel:1991zh}. As this
  time-dependent potential is not important to the exploration of the
  oscillaton structure, owing to the fact that the \eqref{eq:NV2} is decoupled
  from the SP system, its existence was not explored further or even mentioned
  in the most of the literature. On the other hand, this pressure is important
  for understanding the values of the metric coefficients, and ipso facto for
  understanding the motion of test particles in this background.}\BibitemShut
  {Stop}%
\bibitem [{2ca()}]{2caveat}%
  \BibitemOpen
  \href@noop {} {}\bibinfo {note} {These measurements can be degenerate. For
  example, there can be astrophysical sources which could spoil the signal:
  system is a binary, presence of an exoplanet etc. These issues can be in
  principle addressed by looking for the visual confirmation of other object as
  well as detection of the signal on multiple objects. Doppler effect induced
  by radial velocity oscillation is more suppressed as discussed in Section
  \ref{sec:darkhalo_motion}.}\BibitemShut {Stop}%
\end{thebibliography}
\end{document}